\documentclass[a4paper,11pt]{article}
\pdfoutput=1 

\usepackage{jheppub} 

\usepackage[T1]{fontenc} 
\usepackage{graphicx}
\graphicspath{{figure/}} 

\usepackage{lipsum}
\usepackage{amsmath}
\usepackage{bm}
\usepackage[normalem]{ulem}
\usepackage{slashed}
\usepackage{epsfig}
\usepackage{epstopdf}
\usepackage{amsfonts}
\usepackage{subfigure}
\usepackage{xcolor}
\usepackage[outermarks]{titlesec}

\usepackage{bbding}
\usepackage{pifont}
\usepackage{wasysym}
\usepackage{amssymb}

\DeclareMathAlphabet\bfcal{OMS}{cmsy}{b}{n}

\newcommand{\nbar}{{\bar n}}

\newcommand{\beq}{\begin{equation}}
\newcommand{\eeq}{\end{equation}}
\newcommand{\bea}{\begin{eqnarray}}
\newcommand{\eea}{\end{eqnarray}}

\newcommand{\nn}{\nonumber}

\newcommand{\beal}{\begin{align}}
\newcommand{\eeal}{\end{align}}
\newcommand{\bspl}{\begin{split}}
\newcommand{\espl}{\end{split}}

\title{Double-Real-Virtual 
and Double-Virtual-Real
Corrections to the Three-Loop Thrust Soft Function}

\author[a]{Wen Chen,}
\author[b,c]{Feng Feng,}
\author[c,d]{Yu Jia,}
\author[e,f]{and Xiaohui Liu}

\affiliation[a]{School of Physics, Zhejiang University, Hangzhou, Zhejiang 310027, China}
\affiliation[b]{China University of Mining and Technology, Beijing 100083, China}
\affiliation[c]{Institute of High Energy Physics and Theoretical Physics Center for Science Facilities, Chinese Academy of Sciences, Beijing 100049, China}
\affiliation[d]{School of Physics, University of Chinese Academy of Sciences,
Beijing 100049, China}

\affiliation[e]{ Center of Advanced Quantum Studies, Department of Physics, Beijing Normal
University, Beijing 100875}
\affiliation[f]{Center for High Energy Physics, Peking University, Beijing 100871, China}

 \emailAdd{chenwenphy@gmail.com}
 \emailAdd{f.feng@outlook.com}
 \emailAdd{jiay@ihep.ac.cn}
 \emailAdd{xiliu@bnu.edu.cn}

\abstract{We compute the ${\cal O}(\alpha_s^3)$ double-real-virtual (RRV) and double-virtual-real (VVR) soft contributions to the thrust/zero-jettiness event shape. The result clears up one of the most stubborn obstacles toward the complete ${\cal O}(\alpha_s^3)$ thrust soft function. The results presented here serve as the key input to realize the next-to-next-to-next-to-leading logarithmic prime  (N${}^3$LL') 
resummation of the thrust event shape. The obtained results also constitute the important ingredients
of the $N$-jettiness-subtraction scheme at next-to-next-to-next-to-leading order (N${}^3$LO). 
}

\begin{document}
\maketitle
\flushbottom


\section{Introduction}~\label{sec:intr}

The thrust distribution $\tau$ in $e^+e^-$ annihilation process is among the fundamental tools to test the theory of strong
interactions at high precision~\cite{Brandt:1964sa,Farhi:1977sg,Kluth:2006bw}. It probes the global geometrical
structure through an inclusive measurement of 
\bea\label{eq:thrust}
T = 1-\tau = {\rm Max}_{\hat{{\bf t}}} \frac{\sum_i |\hat{{\bf t}} \cdot \vec{k}_i| }{\sum_i|\vec{k}_i|} \,, 
\eea 
where ${\hat{{\bf t}}}$ is known as the thrust axis that maximizes the last equation and $\vec{k}_i$ is the momentum of the particles produced in $e^+e^-$ annihilation. Being theoretically clean and feasible to high perturbative order computations, the thrust distribution is particularly suitable for the precise determination of the strong coupling $\alpha_s(M_Z)$~\cite{Gehrmann-DeRidder:2007nzq,Dissertori:2007xa,Becher:2008cf,Dissertori:2009ik,Abbate:2010xh,Proceedings:2011zvx,Abbate:2012jh,Gehrmann:2012sc} and has been frequently
measured at $e^+e^-$ colliders with small experimental uncertainties. The precise extraction of the $\alpha_s$ with unprecedented accuracy will continue to be the major scientific pillar at future $e^+e^-$ facilities~\cite{Monni:2021rcz}.

The thrust distribution $\tau$ probes the kinematic configurations between a perfectly  isotropic multi-particle final state with $\tau$ close to the tail point $1/2$, and the two collimated
back-to-back energy flows (the di-jet limit) where $\tau \to 0$. 
When sufficiently away from the di-jet limit, the thrust distribution is well modeled by the fixed order calculation which has been known to ${\cal O}(\alpha_s^3)$~\cite{Gehrmann-DeRidder:2007nzq, Gehrmann-DeRidder:2014hxk}. 
Near the di-jet limit in which $\tau\ll 1$, resummation is required to secure the perturbative predictions. 

When $\tau \ll 1$, the $\tau$ spectrum can be written in a factorized form~\cite{Collins:1989gx,Berger:2002ig,Berger:2003iw,Schwartz:2007ib,Becher:2008cf} within the soft collinear effective theory (SCET)~\cite{Bauer:2002nz,Bauer:2001yt,Bauer:2001ct,Bauer:2000yr,Beneke:2002ph}, which gives  
\bea 
\frac{d\sigma_{sing.}}{d\tau} 
= \sigma_0 \int \prod_{i= n,{\bar n},s} d\tau_i {\cal J}_n(\tau_n) {\cal J}_{\bar n}(\tau_{\bar n}) 
S(\tau_s)\delta(\tau - \tau_n - \tau_{\bar n} - 
\tau_s)  \,,
\eea 
where 
$\sigma_0$ contains all loop corrections to the matrix $e^+e^- \to q{\bar q}$ and we assume that $q$ is along the light-like direction $n = (1,0,0,1)$ while ${\bar q}$ along $\bar{n} = (1,0,0,-1)$. 
${\cal J}$ is the inclusive jet function~\cite{Fleming:2003gt,Bauer:2003pi} and $S$ is the soft function defined as the vacuum expectation of Wilson lines~\cite{Becher:2008cf}.
The factorization is valid up to ${\cal O}(\tau)$ power suppressed contributions and with this approximation~\footnote{Systematic studies of the power corrections have also been carried out, see for instance~\cite{Beneke:2022obx,Agarwal:2020uxi} }, the thrust measurement in Eq.~(\ref{eq:thrust}) is simplified to
\bea\label{eq:t-lp}
\tau = \sum_k \min\left[ n\cdot k,{\bar n }\cdot k \right] \,.
\eea 

The factorization theorem reproduces correctly the leading logarithmic singular contributions to the $\tau$ distribution such that
\bea 
 \frac{d \sigma_{sing.}}{d \tau}
= \sum_{n=0}^\infty \sum_{k = 1}^{2n+1} \alpha_s^n \,  C^{(n)}_k  \,  \frac{1}{\mu}
{\cal L}_{2n-k}
\left(\frac{\tau}{\mu}\right) \,, 
\eea 
where the logarithms ${\cal L}$ are given by
\bea 
{\cal L}_{2n-k}(\tau) = 
\left(\frac{\ln^{2n - k} \tau }{\tau} \right)_+ \,, \quad \quad 
{\cal L}_{-1}(\tau)  = \delta(\tau) \,,
\eea 
and the $C_k^{(n)}$ are some constants free of $\tau$.

The factorization theorem sets the foundation for precision prediction of the thrust distribution in the resummation region in which $\tau \ll 1$, and is the key ingredient to the precise determination of the $\alpha_s$. 
The logarithmic terms (i.e., $C_k^{(n)}$ with $k\leqslant 2n$) can be calculated through the standard resummation techniques to all orders in $\alpha_s$ for the jet and the soft functions, if the associated cusp $\Gamma_{cusp}$ and the non-cusp anomalous dimensions $\gamma_i$ are known, see Tab.~\ref{tab:ResummationOrder}. 
The cusp anomalous dimension is known analytically to 4-loops~\cite{Henn:2019swt,vonManteuffel:2020vjv} following earlier efforts~\cite{Ruijl:2016pkm,Moch:2018wjh,Henn:2016men,vonManteuffel:2016xki,Henn:2016wlm,Lee:2017mip,Henn:2019rmi,Lee:2019zop,vonManteuffel:2019wbj,Bruser:2019auj}. The thrust distribution is now known at the next-to-next-to-next-to-leading logarithmic (N$^{3}$LL) accuracy~\cite{Becher:2008cf,Abbate:2010xh,Abbate:2012jh}. 
While the $\delta(\tau)$ terms (i.e., $C_{2n+1}^{(n)}$) have to be determined through explicit fixed order calculations of the jet and the soft function order by order in $\alpha_s$ and their accuracy also determines the logarithmic accuracy of resummation as shown in Tab.~\ref{tab:ResummationOrder}. 
\begin{table*}[ht]
    \centering
    \begin{tabular}{| c | c | c c | }
    \hline
    & & \multicolumn{2}{c |}{Anomalous Dimension}  \\
    Logarithmic Order & $\delta(\tau)$ term & $\gamma_i$  & $\Gamma_{cusp}$     \\
    \hline
    LL & 1 & - & 1-loop   \\
    NLL & 1 & 1-loop & 2-loop  \\
    NLL'   & $\alpha_s$ & 1-loop & 2-loop  \\
    NNLL  & $\alpha_s$ & 2-loop & 3-loop  \\
    NNLL'   & $\alpha_s^2$ & 2-loop & 3-loop  \\
    N${}^3$LL  & $\alpha_s^2$ & 3-loop & 4-loop  \\
    N${}^3$LL'   & $\alpha_s^3$ & 3-loop & 4-loop  \\
    N${}^4$LL   & $\alpha_s^3$ & 4-loop & 5-loop  \\
    \hline
    \end{tabular}
    \caption{The definitions for the logarithmic accuracy of the resummation calculation.}
    \label{tab:ResummationOrder}
\end{table*}

To push through the precision to the next level, the complete $3$-loop calculation for the jet and the soft function have to be performed. The calculation of the N$^{3}$LO quark jet function has been carried out in~\cite{Bruser:2018rad} years ago, 
while the soft function is only known to order ${\cal O}(\alpha_s^2)$~\cite{Monni:2011gb,Kelley:2011ng, Boughezal:2015eha,Baranowski:2020xlp}. Recently, as a first non-trivial step toward the three loops, the same-hemisphere three-gluon real-real-real (RRR) contribution to the soft function is obtained analytically~\cite{Baranowski:2022khd}. 

In this manuscript, we present the computation of the double-real-virtual (RRV) and the double-virtual-real (VVR) contributions to the thrust soft function. Since the $2$-loop soft current is known in the literature for a while~\cite{Li:2013lsa,Duhr:2013msa}, it makes the double-virtual-real (VVR) correction to the thrust soft function easy to obtain, and we present the VVR results in the Appendix~\ref{appendix:vvr}. 

The work is organized as follows: 

In Section~\ref{sec:theory}, we introduce the convention we use in this paper. In Section~\ref{sec:calculation}, we review the method to evaluate the integral in Eq.~(\ref{eq:Srrv-integral}). We demonstrate the methodology by reproducing the known double-real contribution to the ${\cal O}(\alpha_s^2)$ soft function in Section~\ref{subsec:nnlo-soft}. We present the ${\cal O}(\alpha_s^3)$ double-real-virtual correction in Section~\ref{sec:result} and we conclude in Section~\ref{sec:end}. 

\section{Theoretical set-ups}~\label{sec:theory}

\begin{figure}[htbp]
    \centering
    \subfigure{\includegraphics[width=0.325\textwidth]{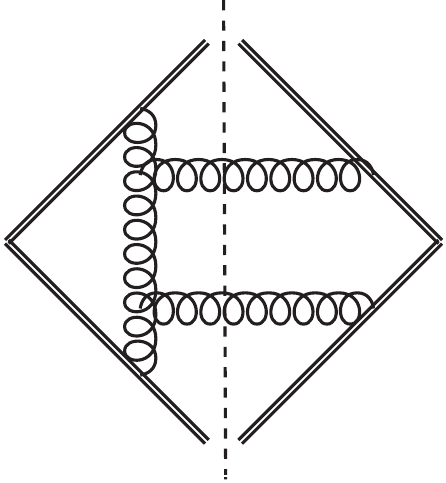}} 
    \subfigure{\includegraphics[width=0.325\textwidth]{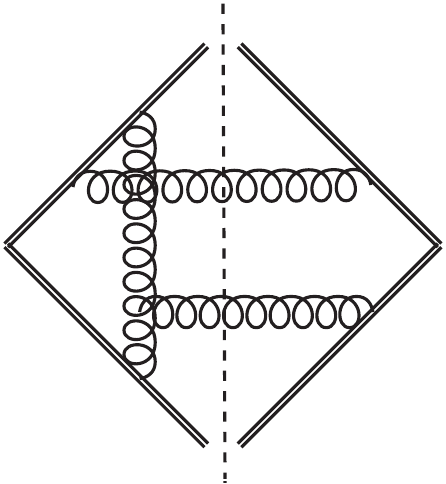}} 
    \subfigure{\includegraphics[width=0.325\textwidth]{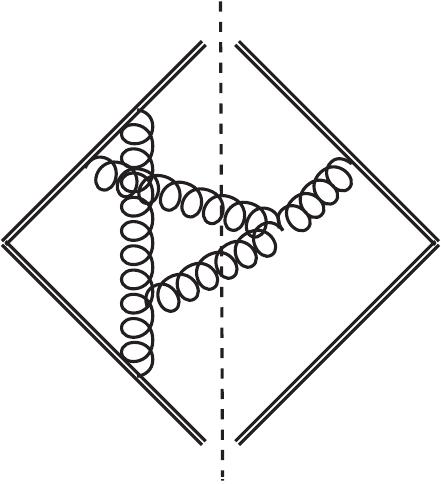}} 
    \subfigure{\includegraphics[width=0.325\textwidth]{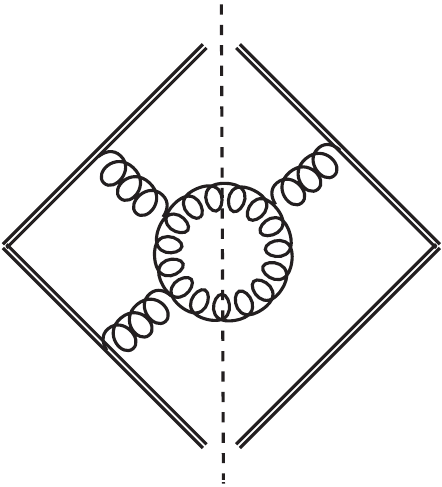}} 
    \subfigure{\includegraphics[width=0.325\textwidth]{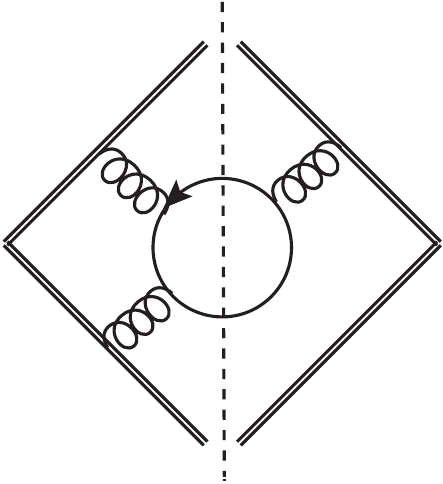}} 
    \subfigure{\includegraphics[width=0.325\textwidth]{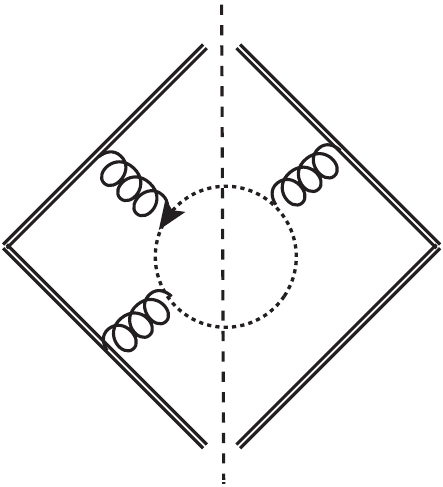}} 
    \subfigure{\includegraphics[width=0.325\textwidth]{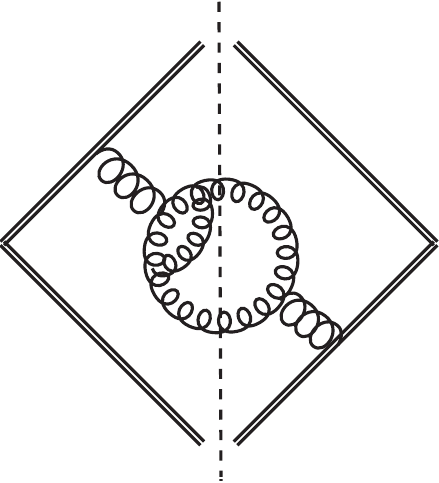}} 
    \subfigure{\includegraphics[width=0.325\textwidth]{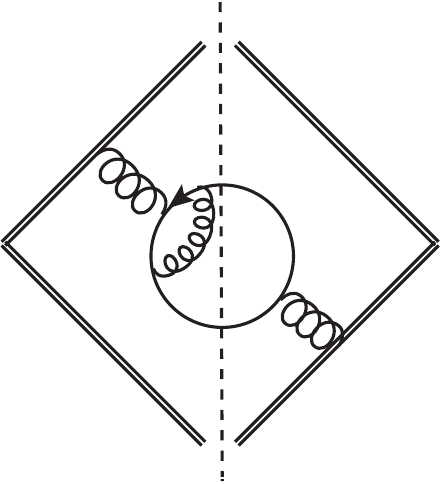}} 
    \subfigure{\includegraphics[width=0.325\textwidth]{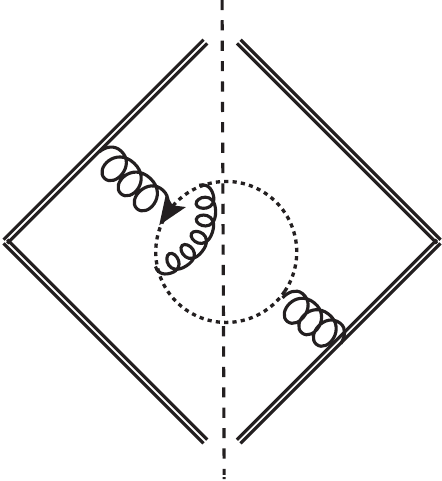}} 
    \caption{Representative Feynman diagrams for the RRV contribution at ${\cal O}(\alpha_s^3)$. Dotted lines represent the ghost propagators while the vertical dashed lines denote the cuts over the real emissions.}
    \label{fig:feyn-rrv}
\end{figure}

We focus on evaluating the RRV corrections,  $S_{RRV}^{(3)}(\tau;\epsilon)$, 
which is given by 
\bea\label{eq:rrv-soft-function} 
S^{(3)}_{RRV}(\tau;\epsilon) =
\int d\Phi_2  \, \Theta(\tau;k_1,k_2)  \,
\int \frac{d^Dl}{(2\pi)^D}\, 
\omega_{RRV}^{(3)}(l,k_1,k_2;n,{\bar n}) \,, 
\eea 
where we used the notation  
\bea 
d\Phi_2 = \frac{d^Dk_1}{(2\pi)^D} 
\frac{d^Dk_2}{(2\pi)^D} (2\pi)\delta^+(k_1^2)
(2\pi) \delta^+(k_2^2) \,,
\eea 
for the $2$ body phase space integration. We have explicitly written out the loop integral $d^Dl$. The matrix element $\omega_{RRV}^{(3)}$ is given by the interference between soft currents with $2$ real emissions at the tree and loop levels, 
\bea 
\omega_{RRV}^{(3)} = 
{\cal M}^{(2),\dagger}_{RR}(k_1,k_2;n,{\bar n})\, 
{\cal M}^{(3)}_{RRV}(l,k_1,k_2;n,{\bar n})
+c.c.. 
\eea 
Representative Feynman diagrams for $\omega_{RRV}^{(3)}$ are shown in fig.~\ref{fig:feyn-rrv}. In our calculation, the Feynman diagrams and amplitudes involving Wilson lines are generated by {\tt Qgraf}~\cite{Nogueira:1991ex}. Additional eikonal propagators and vertexes are included as well as those for conventional QCD in the model. To generate the correct amplitudes, we discard the diagrams containing internal eikonal lines.
Throughout this paper, a $i0^+$ prescription is used for each propagator. That is, the denominator $D_i$ of a propagator should be understood as $D_i+i0^+$. Thus, we write each propagator in $\mathcal{M}^{(L)\dagger}$ as $-\frac{1}{-D_i+i0^+}$. 
The $i 0^+$ prescription is also used for the eikonal propagator, which originated from an outgoing soft gluon emitted from an outgoing energetic parton. 
For brevity, the $i0^+$ will be suppressed hereafter. We note that for the RRV contribution, the final result is independent of the pole position in the eikonal propagator~\cite{Zhu:2020ftr, Catani:2021kcy} and therefore the result applies to the thrust in $e^+e^-$ annihilation, as well as the jettiness distribution in Higgs/Drell-Yan~\cite{Stewart:2010pd, Berger:2010xi} or DIS~\cite{Kang:2012zr,Kang:2013nda}.

The phase space restrictions due to the thrust measure in Eq.~(\ref{eq:t-lp}) are given by
\bea 
\Theta(\tau;k_1,k_2) =  
\int d\tau_1 d\tau_2 
\delta(\tau -\tau_1-\tau_2)\Theta_{\tau_1}(k_1)\Theta_{\tau_2}(k_2) 
\eea 
where we introduced the notation 
\bea 
 \Theta_{\tau}(k) \equiv 
\theta(k^+-k^-)\delta(\tau-k^-)
+\theta(k^- - k^+)\delta(\tau-k^+) \,. 
\eea 

We further introduce the dimensionless variables by re-scaling the momenta as $\tau_i \to \hat{\tau}_i \tau$, $k_i \to \hat{k}_i \tau$ and $l \to \hat{l}  \tau$, to factorize out the $\tau$ dependence from the rest of the integral. We get
\bea\label{eq:Srrv-integral} 
S^{(3)}_{RRV}(\tau;\epsilon) = 
\tau^{-1-6\epsilon} 
\int d\hat{\Phi}_2  \,   \,
 \frac{d^D\hat{l}}{(2\pi)^D}\, 
\omega_{RRV}^{(3)}(\hat{l},\hat{k}_1,\hat{k}_2;n,{\bar n})\, \Theta(1;\hat{k}_1,\hat{k}_2) \,. 
\eea 
In the rest of the work, we will suppress the $\hat{}$ notation. And we will show in detail how the integration in Eq.~(\ref{eq:Srrv-integral}) can be evaluated.

For future convenience, we will always multiply each Dirac $\delta$-function and the Heaviside $\theta$-function by a factor of $2\pi$. We will associate each fold of loop integration with a factor of $\frac{1}{\pi^{D/2}\Gamma(1+\epsilon)}$, where $\epsilon=\frac{1}{2}(4-D)$. 
Throughout the work, we denote $D = 4 - 2 \epsilon $ for the dimensional regularization in $4$-dimension, while we use the notation $d$ for the regularization in a more generic dimensional space-time such as $d = 6-2\epsilon$. 
We relate each fold of integration with respect to a scalar, such as $\int d\tau$, with a factor of $\frac{1}{2\pi}$. That is, instead of calculating $S^{(L)}$, we calculate

\begin{equation}
\hat{S}^{(L)}_{\text{RRV/RR}}(1,\epsilon)\equiv\frac{(2\pi)^3(4\pi)^{LD/2}}{\Gamma(1+\epsilon)^L}S^{(L)}_{\text{RRV/RR}}(1,\epsilon).
\end{equation}

\section{The calculation}\label{sec:calculation}

\subsection{Linear reduction of integrals}

In this part, we briefly review the methodology developed in~\cite{Chen:2019mqc,Chen:2019fzm,Chen:2020wsh} to perform the reduction of the phase-space integrals with kinematic cuts and the conventional Feynman loop integrals in a unified way. An alternative approach was developed in~\cite{Baranowski:2021gxe}. In real calculations, the kinematic cuts are implemented by inserting the Dirac $\delta$-function or the Heaviside $\theta$-functions at the integrand level in the phase space integrals. One example of such kinematic cuts can be seen in Eq.~(\ref{eq:rrv-soft-function}) in the previous section. The unified reduction of the phase space and the Feynman integral roots in the well-known fact that either $\delta$-function or the $\theta$-function has its own integral representation similar to that of a propagator. By virtue of this similarity, integrals with $\delta$-functions and $\theta$-functions can be reduced and calculated by generating linear relations and solving differential equations in the parametric representation.

\subsubsection{Parametric representation}\label{subsubsec:ParaRep}
Here we discuss the parametric representation of the Feynman integrals. To unify the treatment of the phase space integrals and the standard Feynman loop integrals, we note that the $\theta$-function and the $\delta$-function belong to the integral class whose definition is given by
\bea 
w_\lambda(D_i)\equiv e^{-\frac{\lambda+1}{2}i\pi}\int_{-\infty}^{\infty}\mathrm{d}x\frac{1}{x^{\lambda+1}}e^{ix \, D_i} \,. 
\eea 
Obviously, we have
\bea 
w_0(D_i)= 2\pi\theta(D_i)\,,\qquad
w_{-1}(D_i)= 2\pi\delta(D_i)\,, \qquad 
w_{-2}(D_i)= 2\pi\delta^\prime(D_i) \,,
\eea 
for $\lambda = 0$, $\lambda=-1$ and $\lambda=-2$, respectively. 

First, we consider the scalar integrals with the form
\bea \label{eq:DefParInt}
J(\lambda_1,\lambda_2,\dots,\lambda_n)=\int[\mathrm{d}l_1][\mathrm{d}l_2]\cdots[\mathrm{d}l_L]\frac{w_{\lambda_1}(D_1)w_{\lambda_2}(D_2)\cdots w_{\lambda_m}(D_m)}{D_{m+1}^{\lambda_{m+1}+1}D_{m+2}^{\lambda_{m+2}+1}\cdots D_n^{\lambda_n+1}},
\eea 
where $L$ is the number of loops, $l_i$ are the loop momenta, $D_i$ are the denominators of propagators, and we denote the integration measure as $[\mathrm{d}l_i]\equiv\frac{1}{\Gamma(1+\epsilon)}\frac{\mathrm{d}^dl_i}{\pi^{d/2}}$ with $d$ the space-time dimension. 

We start with the all nonnegative $\lambda_i$ case. As was shown in Refs.~\cite{Chen:2019mqc,Chen:2020wsh}, the scalar integral can be parametrized by
\bea\label{eq:param_J} 
J=s_g^{-\frac{L}{2}}e^{i\pi\lambda_f} \, I(\lambda_0,\lambda_1,\ldots,\lambda_n) 
=
s_g^{-\frac{L}{2}}e^{i\pi\lambda_f} \, 
\int \mathrm{d}\Pi^{(n+1)}\mathcal{I}^{(-n-1)}
\,,
\eea 
with $s_g = {\rm det}(\eta_{\mu\nu})$ the determinant of the $d$-dimensional space-time metric, and $\lambda_f=\frac{1}{2}Ld-\frac{1}{2}m-\sum_{i=m+1}^n(\lambda_{i}+1)$. $\lambda_0$ is related to the space-time dimension $d$ through $\lambda_0=-\frac{d}{2}$. The integration measure $\mathrm{d}\Pi^{(n+1)}\equiv \prod_{i=1}^{n+1}\mathrm{d}x_i\delta(1-f(x))$, with the positive definite function $f(x)$ satisfy $f(\alpha x)=\alpha f(x)$. The integrand ${\cal I}^{(-n-1)}$ is a homogeneous function of $x$ of degree $-n-1$ and is given by
\bea \label{eq:ParInt}
\mathcal{I}^{(-n-1)} 
=  \frac{\Gamma(-\lambda_0)}{\Gamma(1+\epsilon)^L\prod_{i=m+1}^{n+1}\Gamma(\lambda_i+1)}
\, \mathcal{F}^{\lambda_0} \, \prod_{i=1}^{n+1}x_i^{\lambda_i}\,, 
\eea 
%
where the polynomial $\mathcal{F}$ is related to the Symanzik polynomials $U$ and $F$ through $\mathcal{F}(x)\equiv F(x)+U(x)x_{n+1}$, with $U(x)\equiv\det{A}$, and $F(x)\equiv U(x)\left(\sum_{i,j=1}^L(A^{-1})_{ij}B_i\cdot B_j-C\right)$. Here $A$, $B$, and $C$ are linear in $x$ and are determined from $\sum_{i=1}^nx_iD_i\equiv\sum_{i,j=1}^LA_{ij}l_i\cdot l_j+2\sum_{i=1}^LB_i\cdot l_i+C$.

The parametrization of the scalar integral $J(\lambda_1,\dots,\lambda_i, \dots, \lambda_n)$ in Eq.~(\ref{eq:param_J}) can be generalized to the cases with negative values of $\lambda_i$ for $i > m $, in which the parametric integral $I$ is defined through
\begin{equation*}
\begin{split}
&I(\lambda_0,\ldots,\lambda_{i-1},-\Lambda,\lambda_{i+1},\ldots,\lambda_n)
\equiv \lim_{\lambda_i\to-\Lambda}I(\lambda_0,\ldots,\lambda_{i-1},\lambda_i,\lambda_{i+1},\ldots,\lambda_n)\\
=&\frac{(-1)^{\Lambda-1}\Gamma(-\lambda_0)}{\Gamma(1+\epsilon)^L\prod_{j=m+1,j\neq i}^{n+1}\Gamma(\lambda_j+1)}\int d\Pi^{(n)}\,
\partial^{\Lambda-1}_{x_i} \mathcal{F}^{\lambda_0}|_{x_i = 0}
\prod_{j\neq i}^{n+1}x_j^{\lambda_j},\quad \Lambda\in N,~i>m.
\end{split}
\end{equation*}
Apparently $I(\lambda_0\,, \dots \,,\lambda_{i-1}\,, -1 \,, \lambda_{i+1}\,, \dots \,, \lambda_n) = I(\lambda_0\,, \dots \,,\lambda_{i-1}\,,   \lambda_{i+1}\,, \dots \,, \lambda_n)$ is understood. 

A tensor integral can also be parametrized in terms of $I(\lambda_0,\dots, \lambda_n)$ recursively in a similar way. That is
\bea~\label{eq:ParTensInt}
J_{i_1i_2\cdots i_r}^{\mu_1\mu_2\cdots\mu_r}&\equiv&\int[\mathrm{d}l_1][\mathrm{d}l_2]\cdots[\mathrm{d}l_L]\frac{w_{\lambda_1}(D_1)w_{\lambda_2}(D_2)\cdots w_{\lambda_m}(D_m)}{D_{m+1}^{\lambda_{m+1}+1}D_{m+2}^{\lambda_{m+2}+1}\cdots D_n^{\lambda_n+1}}l_{i_1}^{\mu_1}l_{i_2}^{\mu_2}\dots l_{i_r}^{\mu_r}\nn \\
&=&s_g^{-L/2}e^{i\pi\lambda_f}\left[P_{i_1}^{\mu_1}P_{i_2}^{\mu_2}\cdots P_{i_r}^{\mu_r}I(\lambda_0,\lambda_1,\lambda_2,\ldots,\lambda_n)\right]_{p^\mu=0}\,,
\eea 
where
\begin{equation}
P_i^\mu(p)\equiv-\frac{\partial}{\partial p_{i,\mu}}-\widetilde{B}_i^\mu(\hat{x})+\frac{1}{2}\sum_{j=1}^L\widetilde{A}_{ij}(\hat{x}) p_j^\mu,
\end{equation}
with $\widetilde{A}_{ij}\equiv {\cal D}_0U(A^{-1})_{ij}$ and $\widetilde{B}_i^\mu\equiv\sum_{j=1}^L\widetilde{A}_{ij}B_j^\mu$. Here $p_i^\mu$ is an auxiliary vector introduced through the identity~\cite{Chen:2019mqc} 
\bea 
l^{\mu_1}_{i_1} \dots 
l^{\mu_r}_{i_r} = \frac{(-1)^r i}{\Gamma(r+1)} \prod_j^r 
\left.\partial_{p_{i_j,\mu_j}} 
\int_0^\infty dy e^{-iy(1+ \sum_i^L p_i \cdot l_i)} \right|_{p_i^\mu = 0} \,,
\eea 
and ${\cal D}_0$ and $\hat{x}_i$ are the index-shifting operators will be given in Eq.~(\ref{eq:operators}) and Eq.~(\ref{eq:xyz}). 

\subsubsection{Linear relations}~\label{subsubsec:LinRel}
The parametric integral satisfies the linear relations
\begin{subequations}\label{eq:IBP_scalar}
\begin{align}
0=&\int \mathrm{d}\Pi^{(n+1)}\frac{\partial}{\partial x_i}\mathcal{I}^{(-n-1)},&& i=1, 2,\ldots, m,\\
0=&\int \mathrm{d}\Pi^{(n+1)}\frac{\partial}{\partial x_i}\mathcal{I}^{(-n-1)}+\delta_{\lambda_i0}\int \mathrm{d}\Pi^{(n)}\left.\mathcal{I}^{(-n)}\right|_{x_i=0},&& i=m+1, m+2,\ldots, n+1.
\end{align}
\end{subequations}
We introduce the index-shifting operators ${\cal R}_i$, ${\cal D}_i$, and ${\cal A}_i$, with $i=0,1,\dots,n$, such that
\bea\label{eq:operators}
{\cal R}_iI(\lambda_0,\dots,\lambda_i,\dots,\lambda_n)&=&(\lambda_i+1)I(\lambda_0,\dots,\lambda_i+1,\dots,\lambda_n),\nn \\
{\cal D}_iI(\lambda_0,\dots,\lambda_i,\dots,\lambda_n)&=&I(\lambda_0,\dots,\lambda_i-1,\dots,\lambda_n),\nn \\
{\cal A}_iI(\lambda_0,\dots,\lambda_i,\dots,\lambda_n)&=&\lambda_iI(\lambda_0,\dots,\lambda_i,\dots,\lambda_n)\,.
\eea 
And we formally define operators ${\cal D}_{n+1}$ and ${\cal R}_{n+1}$, such that ${\cal D}_{n+1}I=I$, and ${\cal R}_{n+1}^iI=({\cal A}_{n+1}+1)({\cal A}_{n+1}+2)\cdots({\cal A}_{n+1}+i)I$, with ${\cal A}_{n+1}\equiv-(L+1){\cal A}_0+\sum_{i=1}^m {\cal A}_i-\sum_{i=m+1}^n({\cal A}_i+1)$. The product of two operators $X$ and $Y$ are defined by $(XY)I\equiv X(YI)$. We further introduce the operators $\hat{x}_i$, $\hat{z}_i$ and $\hat{a}_i$ such that
\bea\label{eq:xyz} 
\hat{x}_i&=&\left\{
\begin{matrix}
{\cal D}_i&,&i=1,~2,\ldots,~m,\\
{\cal R}_i&,&i=m+1,~m+2,\ldots,~n+1,
\end{matrix}
\right.\nn \\
\hat{z}_i&=&\left\{
\begin{matrix}
-{\cal R}_i&,&i=1,~2,\ldots,~m,\\
{\cal D}_i&,&i=m+1,~m+2,\ldots,~n+1,
\end{matrix}
\right.\nn \\
\hat{a}_i&=&\left\{
\begin{matrix}
-{\cal A}_i-1&,&i=1,~2,\ldots,~m,\\
{\cal A}_i&,&i=m+1,~m+2,\ldots,~n+1.
\end{matrix}
\right.
\eea 
Obviously we have $\hat{a}_{n+1}=-(L+1){\cal A}_0-\sum_{i=1}^n(\hat{a}_i+1)$. For $i=1,~2,\ldots,n$, we have the following commutation relations:
%
\bea 
\hat{z}_i\hat{x}_j-\hat{x}_j\hat{z}_i= \delta_{ij},\qquad 
\hat{z}_i\hat{a}_j-\hat{a}_j\hat{z}_i= \delta_{ij}\hat{z}_i,\qquad 
\hat{x}_i\hat{a}_j-\hat{a}_j\hat{x}_i= -\delta_{ij}\hat{x}_i \,.
\eea 
%
With the operators $\hat{x}_i$, $\hat{z}_i$, and $\hat{a}_i$, it is easy to write the IBP identity in Eq.~(\ref{eq:IBP_scalar}) in the following compact form
\begin{equation}\label{eq:ParIBP}
{\cal D}_0\frac{\partial\mathcal{F}(\hat{x})}{\partial \hat{x}_i}-\hat{z}_i = 0,\quad i=1,~2,\dots,n+1.
\end{equation}
One should keep in mind that these operator equations are valid only when they are applied to nontrivial parametric integrals.

\subsubsection{Integral reduction}\label{subsubsec:IntRed}
Similarly to the traditional integration-by-parts (IBP) technique \cite{Tkachov:1981wb,Chetyrkin:1981qh}, parametric integrals can be reduced by solving linear relations generated by Eq.~(\ref{eq:ParIBP}). Based on Eq.~(\ref{eq:ParIBP}), two methods to reduce the parametric integrals were proposed in Ref.~\cite{Chen:2019fzm}. For conventional Feynman integrals, Method II therein is more efficient by virtue of the non-negativity of the indices. However, for integrals with cuts, we have to be able to handle the negative indices, because a delta function is of an index $-1$. So in this case Method I is more advantageous. Here we give a brief introduction to this method.

Let $a_{ij}$ and $b_{ij}$ be the solutions of
\bea 
\sum_jb_{ij}\frac{\partial A(\hat{x})}{\partial \hat{x}_j}=0,\qquad 
\sum_ja_{ij}\frac{\partial B(\hat{x})}{\partial \hat{x}_j}=0,
\eea 
where $A$ and $B$ are defined in the paragraph below Eq.~(\ref{eq:ParInt}). In general, solutions of these two equations could be linearly dependent. We denote the linearly dependent part of the solutions by $c_{ij}$. By default, we assume that these solutions are excluded from $a_{ij}$ and $b_{ij}$ and therefore $a_{ij}$ and $b_{ij}$ are linearly independent. For brevity, we denote
\bea 
\frac{\partial}{\partial a_i}\equiv\sum_ja_{ij}\frac{\partial}{\partial \hat{x}_j},\qquad&& \hat{z}_{a_i}\equiv\sum_ja_{ij}\hat{z}_j,\nn \\
\frac{\partial}{\partial b_i}\equiv\sum_jb_{ij}\frac{\partial}{\partial \hat{x}_j},\qquad&& \hat{z}_{b_i}\equiv\sum_jb_{ij}\hat{z}_j,\nn \\
\frac{\partial}{\partial c_i}\equiv\sum_jc_{ij}\frac{\partial}{\partial \hat{x}_j},\qquad&&
\hat{z}_{c_i}\equiv\sum_jc_{ij}\hat{z}_j.
\eea 
$B_i^\mu$ is of the form $\sum_uB_{iu}Q_u^\mu$, where $Q_u^\mu$ are the linearly independent external momenta. We assume that the Gram determinant $Q_u\cdot Q_v$ is invertible. By introducing certain auxiliary parameters (or, equivalently, by introducing some auxiliary propagators), we can always make the matrices $\frac{\partial B_{j}}{\partial b_i}$ and $\frac{\partial B_{ju}}{\partial b_i}$ invertible. That is, there are matrices $\alpha$ and $\beta$ such that
\bea 
\sum_k\alpha_{ij,k}\frac{\partial A_{mn}}{\partial a_k}=\frac{1}{2}\left(\delta^{im}\delta^{jn}+\delta^{in}\delta^{jm}\right)\,,\qquad 
\sum_k\beta_{iu,k}\frac{\partial B_{jv}}{\partial b_k}= \delta^{ij}\delta^{uv}.
\eea 
Then we have the following identities
\begin{subequations}\label{eq:IBP}
\begin{align}
\frac{\partial C}{\partial c_i}+\hat{z}_{c_i}=&0,\\
\sum_j\bar{B}_{ju}A_{ij}=&B_{iu},\\
\sum_k\bar{A}_{ik}A_{kj}=&\left(A_0+\frac{E}{2}\right)\delta_{ij},
\end{align}
\end{subequations}
where $E$ is the number of external momenta, and
\bea 
\bar{B}_{iu}\equiv \frac{1}{2}\sum_{j,v}g_{uv}\beta_{iv,j}\left(\frac{\partial C}{\partial b_j}+\hat{z}_{b_j}\right)\,, \qquad 
\bar{A}_{ij}\equiv -\bar{B}_i\cdot\bar{B}_j-\sum_k\alpha_{ij,k}\left(\hat{z}_{a_k}+\frac{\partial C}{\partial a_k}\right),
\eea 
where $g_{uv}$ is the inverse of the Gram matrix $Q_u\cdot Q_v$. Comparing with Eqs.~(\ref{eq:ParIBP}), Eqs.~(\ref{eq:IBP}) are of lower degrees in $\hat{x}$. And they are free of ${\cal D}_0$. Thus they will not shift the space-time dimension. In practical calculations, we will use Eqs.~(\ref{eq:IBP}) to generate linear equations between the parametric integrals. The generated equations are then solved by using the package \texttt{Kira}~\cite{Maierhofer:2017gsa,Klappert:2019emp,Klappert:2020aqs,Klappert:2020nbg}.

By using the operators $\bar{B}_i$, Eq.~(\ref{eq:ParTensInt}) can be traded by
\begin{equation}\label{eq:TensGen2}
P_i^\mu=-\frac{\partial}{\partial\bar{p}_{i,\mu}}-\bar{B}_i^\mu+\frac{1}{2}\sum_j\widetilde{A}_{ij}\bar{p}_j^\mu,
\end{equation}
where $\bar{p}_i$ are vectors such that $\bar{p}_i\cdot Q_j=0$. The $\bar{B}$ are free of ${\cal D}_0$, and thus will not shift the spacetime dimension. Due to the definition of $\beta_{ij,k}$, the $\bar{B}$ commute with $\widetilde{A}$'s. Thus by applying operators $P_i^\mu$, tensor integrals are parametrized by integrals of the form $f(\widetilde{A})I(-\frac{d}{2},\ldots)$, where $f(\widetilde{A})$ is a sum of chains of $\widetilde{A}_{ij}$. Chains of $\widetilde{A}_{ij}$ can be traded by sums of chains of $\bar{A}_{ij}$ by solving

\begin{equation}\label{eq:A2Abar}
\begin{split}
\widetilde{A}_{i_2j_2}\widetilde{A}_{i_3j_3}\cdots\widetilde{A}_{i_nj_n}\bar{A}_{i_1j_1}=&\widetilde{A}_{i_1j_1}\widetilde{A}_{i_2j_2}\cdots\widetilde{A}_{i_nj_n}(A_0+\frac{E}{2})\\
&-\frac{1}{2}(\widetilde{A}_{i_1i_2}\widetilde{A}_{j_1j_2}+\widetilde{A}_{i_1j_2}\widetilde{A}_{i_2j_1})\widetilde{A}_{i_3j_3}\cdots\widetilde{A}_{i_nj_n}\\
&-\frac{1}{2}(\widetilde{A}_{i_1i_3}\widetilde{A}_{j_1j_3}+\widetilde{A}_{i_1j_3}\widetilde{A}_{i_3j_1})\widetilde{A}_{i_2j_2}\widetilde{A}_{i_4j_4}\cdots\widetilde{A}_{i_nj_n}\\
&-\cdots\\
&-\frac{1}{2}(\widetilde{A}_{i_1i_n}\widetilde{A}_{j_1j_n}+\widetilde{A}_{i_1j_n}\widetilde{A}_{i_nj_1})\widetilde{A}_{i_2j_2}\widetilde{A}_{i_3j_3}\cdots\widetilde{A}_{i_{n-1}j_{n-1}}.
\end{split}
\end{equation}

\noindent The right-hand side of this equation is of degree $n$ in $\widetilde{A}_{ij}$, while the left-hand side is of degree $n-1$. Thus, by solving these identities, we can reduce the degrees of $\widetilde{A}_{ij}$ recursively.

\subsection{Calculation of master integrals}

After carrying out the linear reduction, we express the amplitudes in terms of linear combinations of master integrals (MIs), which can be evaluated by using the differential-equation method \cite{Kotikov:1990kg,Remiddi:1997ny,Henn:2013pwa}. As was shown in ref. \cite{Chen:2019fzm}, differential equations can be constructed in the parametric representation. We let $y$ be a Lorentz scalar, and assume that $A_{ij}$ are free of $y$. Then we have

\begin{equation}
\frac{\partial}{\partial y}=-\sum_{i,u,v}\bar{B}_{iu}B_{iv}\frac{\partial Q_u\cdot Q_v}{\partial y}-2\sum_{i,u,v}Q_u\cdot Q_v\bar{B}_{iu}\frac{\partial B_{iv}}{\partial y}+\frac{\partial C}{\partial y}.
\end{equation}

As is mentioned in section \ref{sec:intr}, the $\tau$ dependence of soft functions can be factored out. Thus the differentiation of MIs with respect to $\tau$ is trivial. To get a nontrivial differential-equations system (DES), we need to introduce an auxiliary scale. This can be done by inserting into the MIs a trivial integral of the form $\int\mathrm{d}y\delta(K-y)$, where $K$ is a linear combination of Lorentz scalars of the loop momenta. Thus MI $J_i$ becomes
\bea
    J_i\equiv \int[\mathrm{d}k_1][\mathrm{d}k_2][\mathrm{d}l]\mathcal{J}
    =\int\mathrm{d}y\int[\mathrm{d}k_1][\mathrm{d}k_2][\mathrm{d}l]\delta(K-y)\mathcal{J} 
    \equiv \int\frac{\mathrm{d}y}{2\pi}~J^\prime_i(y),
\eea
where the integration measure $[\mathrm{d}l_i]$ is defined below Eq.~(\ref{eq:DefParInt}). The region of integration for $y$ depends on the choice of $K$. The choice of $K$ depends on the integrals to be evaluated. For instance, for integrals free of theta functions (all the theta functions are reduced to delta functions), we insert $\int_0^{1/2}\mathrm{d}y\delta(k_1\cdot k_2-y)$; for integrals containing a theta function $\theta(1-k_1^+-k_2^+)$, we insert $\int_0^{\infty}\mathrm{d}y\delta(k_1^-+k_2^--y)$; etc..

The obtained integrals $J_i(y)$ can be calculated by using the standard differential-equation technique. Different from some papers (see e.g. ref. \cite{Zhu:2014fma}), which regularize the boundary conditions of the differential equations in $6$ dimensions, we calculate $J_i(y)$ in $6$ dimensions from the very beginning. That is $d=D+2=6-2\epsilon$. The dimensional recurrence \cite{Tarasov:1996br} can easily be carried out in the parametric representation. To convert the DES to the canonical form \cite{Henn:2013pwa}, we need to rationalize it first. In the case $K=k_1\cdot k_2-y$, we do the transformation $y=\frac{1}{2}(1-x)^2$. In the case $K=k_1^{+/-}+k_2^{+/-}$, the rationalization is unnecessary, and we do the transformation $y=\frac{1}{x}$. The rationalized DES can be converted to the canonical form by using the package \texttt{epsilon} \cite{Prausa:2017ltv}, which implements Lee's algorithm \cite{Lee:2014ioa,Lee:2020zfb}. After converting the differential-equation system to the canonical form, solving it is straightforward. In the case where $K=k_1\cdot k_2-y$, the singularities of the DES are \{0,~1,~2\}. In the case where $K=k_1^{+/-}+k_2^{+/-}$, the singularities of the DES are \{-1,~0,~1\}. In either case, the solutions of the DES can be expressed in terms of the harmonic polylogarithms \cite{Remiddi:1999ew}, of which some can be converted to normal polylogarithms by using the package \texttt{HPL} \cite{Maitre:2005uu}.

It remains to determine the boundary conditions. We chose the boundary to be at $x=0$. The boundary conditions are determined by matching the asymptotic solutions of the DES to the asymptotic expansions \cite{Beneke:1997zp} of the MIs. Most of the asymptotically expanded integrals can easily be calculated in $d$ dimensions. For those that are not easy to calculate, we determine them through the regularization condition: all $6$-dimensional integrals free of double propagators are free of infrared divergences.

Integrals $J_i$ can be obtained from $J^\prime_i(y)$ by integrating out $y$. For integrals with $K=k_1\cdot k_2-y$, the integration is straightforward. For integrals with $K=k_1^{+/-}+k_2^{+/-}$, $J_i(y)$ is singular at $y=\infty$ ($x=0$). The integration can be carried out by subtracting the singular part of $J(y)$ (denoted by $\text{Sing}\{J(y)\}$) and integrating the singular part in $d$ dimensions. That is,

\begin{equation}
\begin{split}
    J_i=&\int_0^\infty\frac{\mathrm{d}y}{2\pi}~J^\prime_i(y)\\
    =&\int_0^\infty\frac{\mathrm{d}y}{2\pi}~\text{Ser}\{J^\prime_i(y)-\text{Sing}\{J^\prime(y)\}\}+\int_0^\infty\frac{\mathrm{d}y}{2\pi}~\text{Sing}\{J^\prime(y)\}.
\end{split}
\end{equation}

\noindent Here we use $\text{Ser}$ to denote the series expansion of a function with respect to $\epsilon$.

\subsection{An example}\label{subsec:nnlo-soft}
\begin{figure}[htbp]
    \centering
    \includegraphics[width=0.4\textwidth]{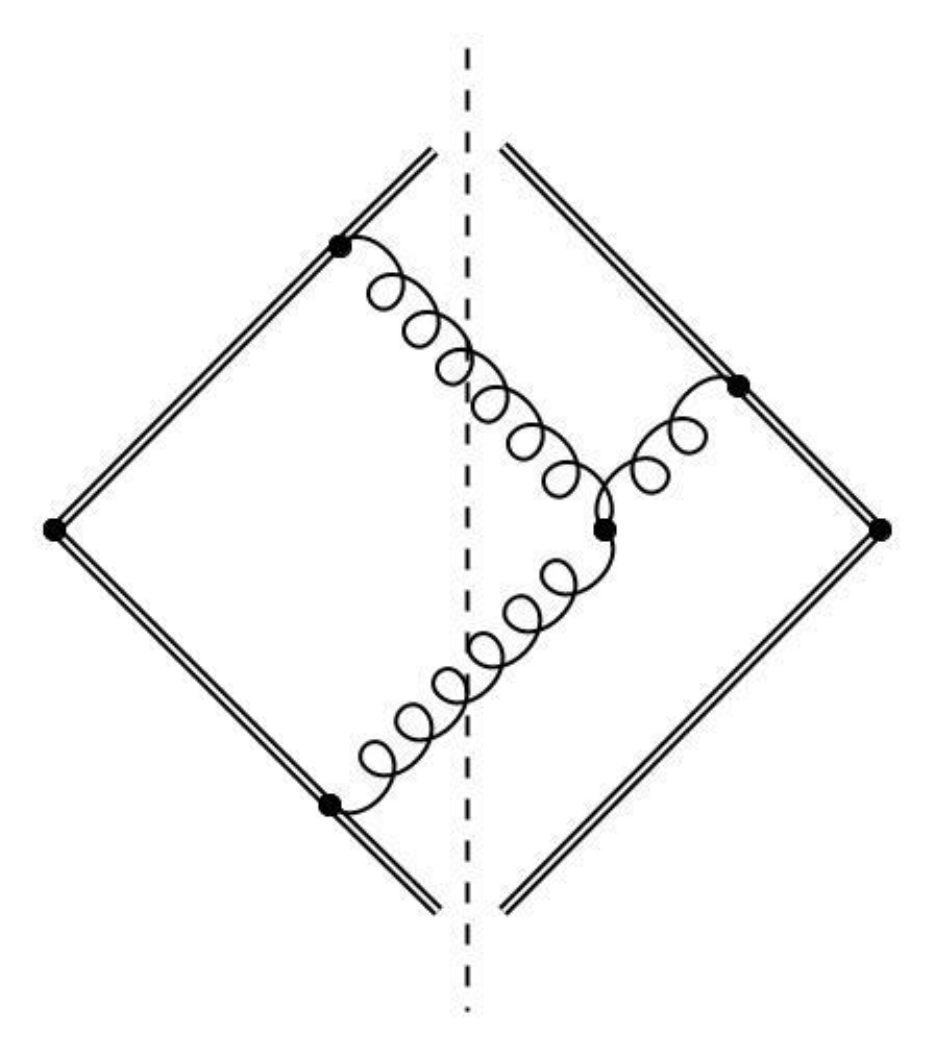}
    \caption{A diagram for the two-gluon emission.}
    \label{fig:feyn-rr}
\end{figure}

To illustrate how to carry out the strategy described in the previous subsections in practice, we present a detailed calculation for the two-gluon-emission diagram shown in Fig. \ref{fig:feyn-rr}. The amplitude for this diagram reads
\bea 
    \hat{S}_{\text{RR,gluon,a}}^{(2)}&=&
    C_AC_Fg_s^4(2\pi)^5\int[\mathrm{d}k_1][\mathrm{d}k_2]\Theta^{(2)}(1,k_1,k_2)\delta(k_1^2)\delta(k_2^2) \nn \\
    && \times\frac{\left(2 k_1^++k_2^+\right)}{k_1^+k_2^-\left(-k_1^+-k_2^+\right) \left[-(k_1+k_2)^2\right]},
\eea
where the colour Casimir constants $C_A=N_C$ and $C_F=\frac{N_C^2-1}{2N_C}$, with $N_C$ the number of colors of a quark. Here we use a subscript $a$ to distinguish it from the full two-gluon emission contribution $\hat{S}_{\text{RR,gluon}}^{(2)}$.

For this amplitude, no tensor reduction is needed. Generally, we use eqs. (\ref{eq:TensGen2},~\ref{eq:A2Abar}) to parametrize tensor integrals. The resulted parametric integrals are reduced by solving linear relations generated by using Eq.~(\ref{eq:IBP}). For example, we consider the following integral family.
%
\bea 
    I(\lambda_0,\lambda_1,\lambda_2,\dots,\lambda_8)&=&i(-1)^{\lambda_6+\lambda_7+\lambda_8}\int [\mathrm{d}k_1] [\mathrm{d}k_2] \nn \\
    &&\hspace{-20.ex}\times \frac{w_{\lambda _1}\left(k_1{}^2\right) w_{\lambda _2}\left(k_1{}^--k_1{}^+\right) w_{\lambda _3}\left(k_2{}^2\right) w_{\lambda _4}\left(k_1{}^++k_2{}^+-1\right) w_{\lambda _5}\left(k_2{}^--k_2{}^+\right)}{(k_1^+)^{\lambda_6+1}(-(k_1+k_2)^2)^{\lambda_7+1}(-k_1^--k_2^-)^{\lambda_8+1}}\,. \quad 
\eea 
%
The operators appearing in Eqs.~(\ref{eq:TensGen2},~\ref{eq:A2Abar}) are
%
\bea 
\bar{B}_1^\mu&=&-\frac{1}{2}(\hat{z}_4+\hat{z}_5-\hat{z}_6+\hat{z}_8-1)n^\mu+\frac{1}{2}\hat{z}_6\bar{n}^\mu,\nn \\
    \bar{B}_2^\mu&=&\frac{1}{2}(\hat{z}_4+\hat{z}_5-\hat{z}_6-1)n^\mu+\frac{1}{2}(\hat{z}_4-\hat{z}_6-1)\bar{n}^\mu, \nn \\
\widetilde{A}&=&\begin{pmatrix}
    \hat{x}_3-\hat{x}_7&\hat{x}_7\\
    \hat{x}_7&\hat{x}_1-\hat{x}_7
    \end{pmatrix},\nn \\
\bar{A}_{11}&=&-\hat{z}_6^2+\hat{z}_4 \hat{z}_6+\hat{z}_5 \hat{z}_6+\hat{z}_8 \hat{z}_6-\hat{z}_6-\hat{z}_1, \nn \\
\bar{A}_{12} &=& \bar{A}_{21}=\frac{1}{2} \hat{z}_4^2+\frac{1}{2} \hat{z}_5 \hat{z}_4-\frac{3}{2} \hat{z}_6 \hat{z}_4+\frac{1}{2} \hat{z}_8 \hat{z}_4-\hat{z}_4+\hat{z}_6^2+\frac{1}{2} \hat{z}_1+\frac{1}{2} \hat{z}_3-\frac{1}{2} \hat{z}_5 \nn \\
&& +\frac{3}{2} \hat{z}_6-\hat{z}_5 \hat{z}_6+\frac{1}{2} \hat{z}_7-\frac{1}{2} \hat{z}_8-\frac{1}{2} \hat{z}_6 \hat{z}_8+\frac{1}{2}, \nn \\
\bar{A}_{22} &=&-\hat{z}_4^2-\hat{z}_5 \hat{z}_4+2 \hat{z}_6 \hat{z}_4+2 \hat{z}_4-\hat{z}_6^2-\hat{z}_3+\hat{z}_5+\hat{z}_5 \hat{z}_6-2 \hat{z}_6-1 \,. 
\eea 
%
It is easy to see that the tensor reduction is carried out by using eqs. (\ref{eq:TensGen2},~\ref{eq:A2Abar}) is consistent with the traditional tensor reduction method. For example, for an integral with a numerator $k_1^-+k_2^-$, it is parametrized by
\bea 
    \bar{n}\cdot(P_1+P_2)I(\lambda_0,\lambda_1,\lambda_2,\dots,\lambda_8)&=&-\bar{n}\cdot(\bar{B}_1+\bar{B}_2)I(\lambda_0,\lambda_1,\lambda_2,\dots,\lambda_8) \nn \\
    &=&\hat{z}_8I(\lambda_0,\lambda_1,\lambda_2,\dots,\lambda_8) \nn \\
    &=&I(\lambda_0,\lambda_1,\lambda_2,\dots,\lambda_8-1) \,.
\eea 
And an integral with a numerator $k_1^2$ is parametrized by
\bea 
    P_1^2I(\lambda_0,\lambda_1,\lambda_2,\dots,\lambda_8)&=&\left[\bar{B}_1^2-\frac{(d-2)}{2}\widetilde{A}_{11}\right]I(\lambda_0,\lambda_1,\lambda_2,\dots,\lambda_8)\nn \\
    &=&\left(\bar{B}_1^2+\bar{A}_{11}\right)I(\lambda_0,\lambda_1,\lambda_2,\dots,\lambda_8)\nn \\
    &=&-\hat{z}_1I(\lambda_0,\lambda_1,\lambda_2,\dots,\lambda_8) \nn \\
    &=&(\lambda_1+1)I(\lambda_0,\lambda_1+1,\lambda_2,\dots,\lambda_8)\,. 
\eea 
The IBP identities for this integral family are given by
\bea 
0&=&\hat{x}_1 \hat{z}_4+\hat{x}_1 \hat{z}_5-\hat{x}_1 \hat{z}_6+\hat{x}_1 \hat{z}_8-\hat{x}_7 \hat{z}_8-\hat{x}_1-\hat{x}_2+\hat{x}_4+\hat{x}_6,\nn\\
0&=&\hat{x}_3 \left(-\hat{z}_4\right)-\hat{x}_3 \hat{z}_5+\hat{x}_3 \hat{z}_6-\hat{x}_7 \hat{z}_8+\hat{x}_3+\hat{x}_4-\hat{x}_5,\nn\\
0&=&\hat{x}_7 \hat{z}_4-\hat{x}_1 \hat{z}_6+\hat{x}_2-\hat{x}_7-\hat{x}_8,\nn\\
0&=&\hat{x}_3 \left(-\hat{z}_4\right)+\hat{x}_3 \hat{z}_6+\hat{x}_7 \hat{z}_4+\hat{x}_3+\hat{x}_5-\hat{x}_7-\hat{x}_8,\nn\\
0&=&2 a_0+2 a_1+a_6+a_7+a_8-\hat{x}_7 \hat{z}_1+\hat{x}_7 \hat{z}_3-\hat{x}_2 \hat{z}_4+\hat{x}_8 \hat{z}_4-\hat{x}_2 \hat{z}_5+\hat{x}_8 \hat{z}_5+\hat{x}_4 \hat{z}_6\nn\\
&&-\hat{x}_8 \hat{z}_6-\hat{x}_2\hat{z}_8+\hat{x}_2-\hat{x}_8+5,\\
0&=&-a_3-a_5+a_7+a_8-\hat{x}_3 \hat{z}_1-\hat{x}_7 \hat{z}_1+\hat{x}_7 \hat{z}_3-\hat{x}_5 \hat{z}_4+\hat{x}_8 \hat{z}_4+\hat{x}_8 \hat{z}_5+\hat{x}_4 \hat{z}_6-\hat{x}_8 \hat{z}_6\nn\\
&&-\hat{x}_3 \hat{z}_7-\hat{x}_5 \hat{z}_8+\hat{x}_5-\hat{x}_8,\nn\\
0&=&-a_1+a_4-a_6+a_7+\hat{x}_7 \hat{z}_1-\hat{x}_1 \hat{z}_3-\hat{x}_7 \hat{z}_3+\hat{x}_6 \hat{z}_4-\hat{x}_8 \hat{z}_4+\hat{x}_2 \hat{z}_5-\hat{x}_8 \hat{z}_5-\hat{x}_4 \hat{z}_6\nn\\
&&+\hat{x}_8 \hat{z}_6-\hat{x}_1 \hat{z}_7-\hat{x}_4-\hat{x}_6+\hat{x}_8,\nn\\
0&=&2 a_0+2 a_3+a_4+a_5+a_7+\hat{x}_7 \hat{z}_1-\hat{x}_7 \hat{z}_3-\hat{x}_8 \hat{z}_4-\hat{x}_8 \hat{z}_5-\hat{x}_4 \hat{z}_6+\hat{x}_8 \hat{z}_6-\hat{x}_4+\hat{x}_8+5,\nn\\
0&=&\hat{z}_2+\hat{z}_4+\hat{z}_5+\hat{z}_8-1.\nn
\eea 
After carrying out the IBP reduction we further express the resulted amplitude in terms of $6$-dimensional integrals through dimensional recurrence:
\bea 
    I(\lambda_0,\lambda_1,\lambda_2,\dots,\lambda_8)&=&{\cal D}_0U~I(\lambda_0,\lambda_1,\lambda_2,\dots,\lambda_8)\nn \\
    &=&\left(\hat{x}_3 \hat{x}_1-\hat{x}_7 \hat{x}_1-\hat{x}_7 \hat{x}_3\right)I(\lambda_0-1,\lambda_1,\lambda_2,\dots,\lambda_8).
\eea 
The reductions for all the other families of integrals are similar. We finally find 
\bea 
    \hat{S}_{\text{RR,gluon,a}}^{(2)}&=&C_AC_Fg_s^4\times\nn \\
    &&\left[\frac{\left(118240 \epsilon ^7-469064 \epsilon ^6+765604 \epsilon ^5-677718 \epsilon ^4+356451 \epsilon ^3-112281 \epsilon ^2+19542 \epsilon -1440\right)}{6 \epsilon ^3 (\epsilon -1) (2 \epsilon -1) (3 \epsilon -2) (3 \epsilon -1)}J_1\right. \nn \\
    &&-\frac{4 (\epsilon -1) (4 \epsilon -5) (4 \epsilon -3) \left(37 \epsilon ^2-27 \epsilon +5\right)}{9 \epsilon ^2 (3 \epsilon -2) (3 \epsilon -1)} J_2+\frac{5 (\epsilon -1) (4 \epsilon -3) (4 \epsilon -1)}{\epsilon ^2} J_3 \nn \\
    &&-\frac{(\epsilon -1) (2 \epsilon -1)}{\epsilon }J_4 +\frac{2 (\epsilon -1) (4 \epsilon -5) \left(256 \epsilon ^3-336 \epsilon ^2+132 \epsilon -15\right)}{\epsilon ^2 (2 \epsilon -1)} J_5 \nn \\
    &&
    +\frac{4 (\epsilon -1) (4 \epsilon -5) (4 \epsilon -3) \left(230 \epsilon ^3-193 \epsilon ^2+55 \epsilon -5\right)}{9 \epsilon ^2 (2 \epsilon -1) (3 \epsilon -2) (3 \epsilon -1)} J_6-\frac{4 (\epsilon -1) (4 \epsilon -3) (4 \epsilon -1)}{\epsilon } J_7 \nn \\
    &&
    \left.+\frac{4 (\epsilon -1) (4 \epsilon -3) (4 \epsilon -1)}{\epsilon } J_8-\frac{ (\epsilon -1) (2 \epsilon -1)}{\epsilon }J_9\right]. 
\eea 
Here the MIs $J_i$ are give by~\footnote{According to the $i0^+$ prescription of the denominator of a Feynman propagator, an integral with a denominator, say, $k_1^-+k_2^-$, will be treated as independent of the integral with a denominator $-k_1^--k_2^-$.}

\bea 
J_1 &=& (2\pi)^5\int [\mathrm{d}k_1] [\mathrm{d}k_2] \delta \left(k_1^2\right) \delta \left(k_2^2\right) \delta \left(k_1^++k_2^+-1\right) \delta \left(k_1^--k_1^+\right) \delta \left(k_2^--k_2^+\right),\nn \\
J_2 &=& (2\pi)^5\int [\mathrm{d}k_1] [\mathrm{d}k_2]\frac{ \delta \left(k_1^2\right) \delta \left(k_2^2\right) \delta \left(k_1^++k_2^+-1\right) \delta \left(k_2^--k_2^+\right) \theta \left(k_1^--k_1^+\right)}{-k_1^--k_2^-},\nn\\
J_3 &=& (2\pi)^5\int [\mathrm{d}k_1] [\mathrm{d}k_2]\frac{ \delta \left(k_1^2\right) \delta \left(k_2^2\right) \delta \left(k_1^++k_2^+-1\right) \delta \left(k_2^--k_2^+\right) \theta \left(k_1^--k_1^+\right)}{k_1^- \left[-(k_1+k_2)^2\right]},\nn\\
J_4 &=& (2\pi)^5\int [\mathrm{d}k_1] [\mathrm{d}k_2]\frac{ \delta \left(k_1^2\right) \delta \left(k_2^2\right) \delta \left(k_1^++k_2^+-1\right) \delta \left(k_2^--k_2^+\right) \theta \left(k_1^--k_1^+\right)}{k_1^+ \left(-k_1^--k_2^-\right) \left[-(k_1+k_2)^2\right]},\nn\\
J_5 &=& (2\pi)^5\int [\mathrm{d}k_1] [\mathrm{d}k_2]\frac{ \delta \left(k_1^2\right) \delta \left(k_2^2\right) \delta \left(k_1^++k_2^--1\right) \theta \left(k_1^--k_1^+\right) \theta \left(k_2^+-k_2^-\right)}{-(k_1+k_2)^2},\nn\\
J_6 &=& (2\pi)^5\int [\mathrm{d}k_1] [\mathrm{d}k_2]\frac{ \delta \left(k_1^2\right) \delta \left(k_2^2\right) \delta \left(k_1^++k_2^--1\right) \delta \left(k_1^--k_1^+\right) \theta \left(k_2^+-k_2^-\right)}{-k_1^+-k_2^+},\nn\\
J_7 &=& (2\pi)^5\int [\mathrm{d}k_1] [\mathrm{d}k_2]\frac{ \delta \left(k_1^2\right) \delta \left(k_2^2\right) \delta \left(k_1^++k_2^--1\right) \theta \left(k_1^--k_1^+\right) \theta \left(k_2^+-k_2^-\right)}{k_1^- \left(-k_1^+-k_2^+\right) \left[-(k_1+k_2)^2\right]},\nn\\
J_8 &=& (2\pi)^5\int [\mathrm{d}k_1] [\mathrm{d}k_2]\frac{ \delta \left(k_1^2\right) \delta \left(k_2^2\right) \delta \left(k_1^++k_2^--1\right) \theta \left(k_1^--k_1^+\right) \theta \left(k_2^+-k_2^-\right)}{k_1^- k_2^+ \left[-(k_1+k_2)^2\right]},\nn\\
J_9 &=& (2\pi)^5\int [\mathrm{d}k_1] [\mathrm{d}k_2]\frac{ \delta \left(k_1^2\right) \delta \left(k_2^2\right) \delta \left(k_1^++k_2^--1\right) \delta \left(k_1^--k_1^+\right) \theta \left(k_2^+-k_2^-\right)}{\left(-k_1^+-k_2^+\right) k_2^- \left[-(k_1+k_2)^2\right]}.
\eea 
For the readers' convenience, we have expressed the parametric integrals in terms of the normal momentum-space integrals. These MIs $J_i$ can be divided into two classes. The measure of one class includes a delta function $\delta(k_1^++k_2^+-1)$. And the that of another class includes a delta function $\delta(k_1^++k_2^--1)$.~\footnote{For this calculation, all the integrals with delta functions $\delta(k_1^-+k_2^--1)$ and $\delta(k_1^-+k_2^+-1)$ can be identified with the integrals with delta functions $\delta(k_1^++k_2^+-1)$ and $\delta(k_1^++k_2^--1)$, by virtue of the symmetries under permutations of indices in the parametric representation.} Here we consider the first class. That is, integrals $J_{1-4}$.

The first integral $J_1$ is free of theta functions. We insert to it a delta function $\delta(k_1\cdot k_2-y)$, and replace $y$ by $y=\frac{1}{2}(1-x)^2$. As in the convention used in the last subsection, we use $J^\prime_i(y)$ to denote an integral resulting from inserting a delta function into $J_i$. The differential equation for $J^\prime_1$ is simple:
\bea 
    \frac{dJ^\prime_1(x)}{dx}=\left(\frac{1-2 \epsilon }{x-1}+\frac{1-\epsilon }{x-2}+\frac{1-\epsilon }{x}\right)J^\prime_1(x),
\eea 
which can be solved easily. The boundary condition can also be readily determined.

For the integrals $J_{2,3,4}$, we insert a delta function $\delta(k_1^-+k_2^--y)$. The resulting integrals $J^\prime_2(y)$ and $J^\prime_4(y)$ can further be reduced. We have
\bea 
    J^\prime_2(y) &=& -\frac{1}{y}(2\pi)^6\int [\mathrm{d}k_1] [\mathrm{d}k_2] \delta \left(k_1^2\right) \delta \left(k_2^2\right) \delta \left(k_1^++k_2^+-1\right) \delta \left(k_2^--k_2^+\right) \theta \left(k_1^--k_1^+\right) \delta \left(-y+k_1^-+k_2^-\right)\nn \\
    &\equiv&-\frac{1}{y}J^\prime_{10}(y), \nn \\
J^\prime_4(y) &=& -\frac{1}{y}(2\pi)^6\int [\mathrm{d}k_1] [\mathrm{d}k_2]\frac{ \delta \left(k_1^2\right) \delta \left(k_2^2\right) \delta \left(k_1^++k_2^+-1\right) \delta \left(k_2^--k_2^+\right) \theta \left(k_1^--k_1^+\right) \delta \left(-y+k_1^-+k_2^-\right)}{k_1^+ \left(-(k_1+k_2)^2\right)}\nn \\
&\equiv&-\frac{1}{y}J^\prime_{11}(y).
\eea 
To get a closed DES, we need to add one more integral
\bea 
J^\prime_{12}(y) = (2\pi)^6\int [\mathrm{d} k_1] [\mathrm{d} k_2] \delta\left(k_1^2\right) \delta \left(k_2^2\right) \delta \left(k_1^++k_2^+-1\right) \delta \left(k_2^--k_2^+\right) \theta \left(k_1^--k_1^+\right) \delta^\prime \left(-y+k_1^-+k_2^-\right).\qquad
\eea 
The DES for the integrals $\{J^\prime_{12},~J^\prime_{10},~J^\prime_{2},~J^\prime_{11}\}$ is
\bea 
    \frac{d}{dy}\begin{pmatrix}J^\prime_{12}\\J^\prime_{10}\\J^\prime_{3}\\J^\prime_{11}\end{pmatrix}=
    \begin{pmatrix}
 -\frac{(5 y-3) (\epsilon -1)}{(y-1) y} & \frac{(\epsilon -1) (4 \
\epsilon -5)}{(y-1) y} & 0 & 0 \\
 -1 & 0 & 0 & 0 \\
 \frac{3 y^2-2 y+3}{2 (y-1)^2 y^2} & -\frac{(3 y-1) (4 \epsilon \
-5)}{2 (y-1)^2 y^2} & -\frac{(2 y-1) (2 \epsilon -1)}{(y-1) y} & 0 \\
 \frac{3 y^2-2 y+3}{2 (y-1)^2 y} & -\frac{(3 y-1) (4 \epsilon -5)}{2 \
(y-1)^2 y} & 0 & -\frac{2 \epsilon -1}{y-1} 
    \end{pmatrix}.
    \begin{pmatrix}J^\prime_{12}\\J^\prime_{10}\\J^\prime_{3}\\J^\prime_{11}\end{pmatrix}\,,
\eea 
which can easily be transformed to the canonical form by using \texttt{epsilon}~\cite{Prausa:2017ltv}.

To determine the boundary conditions, we change the variable to $x=\frac{1}{y}$. The asymptotic solutions of the DES are of the form
\bea 
J^\prime_i(x)\approx \sum_j \left(C_{1,i,j}x^{\epsilon+j}+C_{2,i,j}x^{2\epsilon+j}+C_{3,i,j}x^{4\epsilon+j}\right).
\eea 
Among all the $C_{1,i,j}$, only $C_{1,10,-1}$ is independent, which can be determined by calculating the asymptotic expansions of $J^\prime_{10}$. And we note that all the $C_{2,i,j}$ and $C_{3,i,j}$ vanish. To see this, we rescale all the loop momenta by a factor of $x$. For example,
\bea    
J^\prime_{10}(x)&=&(2\pi)^6x^{4\epsilon-5}\int [\mathrm{d}k_1] [\mathrm{d}k_2] \delta \left(k_1^2\right) \delta \left(k_2^2\right) \delta \left(k_1^++k_2^+-x\right) \nn \\ 
    && \hspace{5.3ex}
    \times \delta \left(k_2^--k_2^+\right) \theta \left(k_1^--k_1^+\right) \delta \left(-1+k_1^-+k_2^-\right).
\eea 
Due to the constraints of the delta functions, it is easy to see that there is only one region, in which $k_1^+\sim k_2^+\sim k_2^-\sim x$, and $k_1^-\sim 1$. Thus $J^\prime_{10}(x)\sim x^{\epsilon-1}$, so only $C_{1,i,j}$ survive. Once the boundary conditions are determined, solving the DES is straightforward.

It remains to calculate $J_i$ by integrating $J^\prime_i(y)$ with respect to $y$ (or $x$). Obviously, the integration is divergent when $y\to\infty$ (or $x\to0$). To carry out the integration, we need to subtract $J^\prime_i(y)$ by their singular parts and integrate the singular parts in $d$ dimensions. The singular parts are nothing but the asymptotic solutions we obtained before.

Finally, we solved all the MIs,
\bea 
\frac{J_1}{(2\pi)^3}=&\epsilon ^3 \left(-\frac{77}{75} \zeta_2-\frac{8 \zeta_3}{15}+\frac{277013}{101250}\right)+\epsilon ^2 \left(\frac{4699}{6750}-\frac{\zeta_2}{5}\right)+\frac{77 \epsilon }{450}+\frac{1}{30},\nn \\
\frac{J_2}{(2\pi)^3}=&\epsilon ^2 \left(-\frac{19}{25} \zeta_2-\frac{\zeta_3}{5}+\frac{81061}{33750}\right)+\epsilon  \left(\frac{14783}{18000}-\frac{3 \zeta_2}{20}\right)+\frac{1}{20 \epsilon }+\frac{97}{400},\nn \\
\frac{J_3}{(2\pi)^3}=&\epsilon ^2 \left(\frac{77 \zeta_2}{18}+\frac{7 \zeta_3}{3}-\frac{1055}{81}\right)+\epsilon  \left(\frac{5 \zeta_2}{6}-\frac{95}{27}\right)-\frac{1}{6 \epsilon }-\frac{8}{9},\nn \\
\frac{J_4}{(2\pi)^3}=&\epsilon  \left(-\frac{3}{2} \zeta_2+2 \zeta_3+\frac{19}{4}\right)+\frac{1}{2 \epsilon }+\frac{9}{4},\nn \\
\frac{J_5}{(2\pi)^3}=&\epsilon ^2 \left(\frac{29 \zeta_2}{45}+\frac{7 \zeta_3}{30}-\frac{126722}{50625}\right)+\epsilon  \left(\frac{\zeta_2}{12}-\frac{16523}{27000}\right)-\frac{1}{60 \epsilon }-\frac{229}{1800},\nn \\
\frac{J_6}{(2\pi)^3}=&\epsilon ^2 \left(-\frac{19}{25} \zeta_2-\frac{\zeta_3}{5}+\frac{81061}{33750}\right)+\epsilon  \left(\frac{14783}{18000}-\frac{3 \zeta_2}{20}\right)+\frac{1}{20 \epsilon }+\frac{97}{400},\nn \\
\frac{J_7}{(2\pi)^3}=&-\frac{1}{2} \zeta_2+\epsilon  \left(-\frac{71}{36} \zeta_2-\frac{\zeta_3}{3}+\frac{1715}{324}\right)+\frac{3}{4 \epsilon }+\frac{1}{6 \epsilon ^2}+\frac{247}{108},\nn \\
\frac{J_8}{(2\pi)^3}=&\zeta_2+\epsilon  \left(\frac{11 \zeta_2}{2}+\frac{5 \zeta_3}{2}-\frac{2389}{162}\right)-\frac{8}{9 \epsilon }-\frac{1}{6 \epsilon ^2}-\frac{199}{54},\nn \\
\frac{J_9}{(2\pi)^3}=&\epsilon  \left(-\frac{3}{2} \zeta_2+2 \zeta_3+\frac{19}{4}\right)+\frac{1}{2 \epsilon }+\frac{9}{4}.
\eea 
Here $\zeta_n$ is the Riemann $\zeta$-function. Gathering all the pieces, we find 
\bea 
    \hat{S}_{\text{RR,gluon,a}}^{(2)}=(2\pi)^3C_AC_Fg_s^4\left(\frac{6}{\epsilon}\zeta_2+26\zeta_3\right).
\eea 

For completeness, we present the complete double-real contributions (up to ${\cal O}(\epsilon^2)$) in Appendix \ref{appendix:rr}. The results are in full agreement with the known calculations~\cite{Monni:2011gb,Kelley:2011ng, Boughezal:2015eha,Baranowski:2020xlp}, which serves as a strong validation of our computational framework.

\section{Results}~\label{sec:result}
Now we present the result for RRV. 
We start with the $2$ gluon emission contribution. In the virtual loop, all parton propagators are included. 
We decompose the two-gluon-virtual contribution according to the color structures by\footnote{Here the factor $\cos(\pi\epsilon)$ arises from the phase factor $e^{i\pi\epsilon}$ of the amplitude, which is absent when one of the Wilson lines is incoming and the other is outgoing.}

\begin{equation}\label{eq:srrvg}
\begin{split}
    \hat{S}^{(3)}_{\mathrm{RRV,} gg}=&(4\pi)^6\cos(\pi\epsilon)\alpha_s^3C_AC_F\\
    &\times\left(N_f\hat{S}^{(3)}_{\mathrm{RRV,} gg,a}+C_A\hat{S}^{(3)}_{\mathrm{RRV,} gg,b}+C_F\hat{S}^{(3)}_{\mathrm{RRV,} gg,c}\right),
\end{split}
\end{equation}

\noindent where $N_f$ is the number of the quark flavors, and $C_A$ and $C_F$ are color structures defined at the beginning of section \ref{subsec:nnlo-soft}. Here

\begin{equation}
    \begin{split}
        \hat{S}^{(3)}_{\mathrm{RR,} gg,a}=&\frac{1}{54 \epsilon ^3}+\frac{31}{162 \epsilon ^2}+\frac{1}{\epsilon }\left(\frac{95}{81}-\frac{1}{9}\zeta_2\right)+\frac{17}{27} \zeta_2-\frac{41}{27} \zeta_3+\frac{3778}{729}\\
        &+\epsilon\left[\frac{62}{27} \zeta_2+\frac{43}{81} \zeta_3-\frac{149}{18} \zeta_4+\frac{39568}{2187}\right],
    \end{split}
\end{equation}

\begin{equation}
    \begin{split}
        \hat{S}^{(3)}_{\mathrm{RRV,} gg,b}=&-\frac{5}{3 \epsilon ^5}-\frac{523}{144 \epsilon ^4}+\frac{1}{\epsilon ^3}\left(\frac{34 }{3}\zeta_2-\frac{2129}{216}\right)+\frac{1}{\epsilon ^2}\left(\frac{409}{24} \zeta_2+42 \zeta_3-\frac{76}{3}\right)\\
        &+\frac{1}{\epsilon}\left(-\frac{386}{3} \zeta_2+\frac{14399 }{72}\zeta_3+\frac{478}{3} \zeta_4-\frac{101329}{1944}\right)-\frac{596}{3} \zeta_3 \zeta_2\\
        &+\frac{6277}{54} \zeta_2-\frac{121919 }{108}\zeta_3+\frac{8889}{8} \zeta_4+\frac{2360}{3} \zeta_5-4 \log (2) \zeta_2-\frac{649801}{5832}\\
        &+\epsilon\left[48 \mathrm{Li}_4\left(\frac{1}{2}\right)+\frac{2180}{3} \zeta _3^2-\frac{95291}{24} \zeta _2 \zeta _3+\frac{30373}{27} \zeta _3-\frac{126233}{162} \zeta _2-\frac{435631}{72} \zeta _4\right.\\
        &\left.+\frac{216925}{16} \zeta _5+\frac{4951}{9} \zeta _6-8 \zeta _2 \log ^2(2)+\frac{59}{3} \zeta _2 \log (2)-\frac{795259}{5832}+2 \log ^4(2)\right],
    \end{split}
\end{equation}
and 
\begin{equation}
    \begin{split}
        \hat{S}^{(3)}_{\mathrm{RRV,} gg,c}=&\frac{6}{\epsilon ^5}-\frac{60}{\epsilon ^3} \zeta_2-\frac{300}{\epsilon ^2} \zeta_3-\frac{810}{\epsilon} \zeta_4+3000 \zeta_2 \zeta_3-8100 \zeta_5\\
        &+\epsilon\left[7500 \zeta_3^2-19635 \zeta_6\right].
    \end{split}
\end{equation}
Similarly, for the two-ghost-virtual contribution and the fermionic contribution, we organize the contribution according to the color decomposition as  
\begin{align}    \hat{S}^{(3)}_{\mathrm{RRV,gh.gh.}}=&2(4\pi)^6\cos(\pi\epsilon)\alpha_s^3C_AC_F\left(N_f\hat{S}^{(3)}_{\mathrm{RRV,gh.gh.},a}+C_A\hat{S}^{(3)}_{\mathrm{RRV,gh.gh.},b}\right),\\
\hat{S}^{(3)}_{\mathrm{RRV,}q\bar{q}}=&2(4\pi)^6\cos(\pi\epsilon)\alpha_s^3N_fC_F\left(N_f\hat{S}^{(3)}_{\mathrm{RRV,}q\bar{q},a}+C_A\hat{S}^{(3)}_{\mathrm{RRV,}q\bar{q},b}+C_F\hat{S}^{(3)}_{\mathrm{RRV,}q\bar{q},c}\right),
\end{align}

\noindent with

\begin{equation}
    \begin{split}
        \hat{S}^{(3)}_{\mathrm{RRV,gh.gh.},a}=&-\frac{1}{108 \epsilon ^3}-\frac{13}{324 \epsilon ^2}+\frac{1}{\epsilon}\left(\frac{1}{18}\zeta_2-\frac{10}{81}\right)-\frac{17}{54} \zeta_2+\frac{41}{54} \zeta_3-\frac{566}{729}\\
        &+\epsilon\left[\frac{26}{27} \zeta_2-\frac{457}{162} \zeta_3+\frac{149}{36} \zeta_4-\frac{14575}{4374}\right],
    \end{split}
\end{equation}
\begin{equation}
    \begin{split}
        \hat{S}^{(3)}_{\mathrm{RRV,gh.gh.},b}=&-\frac{5}{288 \epsilon ^4}-\frac{5}{144 \epsilon ^3}+\frac{1}{\epsilon ^2}\left(\frac{5}{144} \zeta_2-\frac{2}{81}\right)+\frac{1}{\epsilon }\left(-\frac{5}{27} \zeta_2+\frac{11}{48} \zeta_3-\frac{167}{3888}\right)\\
        &-\frac{55}{324} \zeta_2-\frac{289}{216} \zeta_3-\frac{1}{16}\zeta_4+2 \log (2) \zeta_2+\frac{35}{48}\\
        &+\epsilon\left[\frac{6583 \zeta_2}{972}-\frac{655}{144} \zeta_2 \zeta_3-\frac{529}{162} \zeta_3+\frac{2543}{144} \zeta_4+\frac{425}{96} \zeta_5+4 \log^2(2) \zeta_2\right.\\
        &\left.-\frac{59}{6} \log (2) \zeta_2-24 \mathrm{Li}_4\left(\frac{1}{2}\right)+\frac{20513}{34992}-\log^4 (2)\right],
    \end{split}
\end{equation}
and the quark contribution gives 
\begin{equation}
    \begin{split}
        \hat{S}^{(3)}_{\mathrm{RRV,} q\bar{q},a}=&\frac{2}{27 \epsilon ^3}+\frac{20}{81 \epsilon ^2} +\frac{1}{\epsilon }\left(\frac{2}{3}-\frac{4}{9} \zeta _2\right)+\frac{116}{27} \zeta _2-\frac{164}{27} \zeta _3 -\frac{80}{729}\\
        &+\epsilon\left(-\frac{52}{9} \zeta _2+\frac{3508}{81} \zeta _3-\frac{298}{9} \zeta _4-\frac{13118}{2187}\right) ,
    \end{split}
\end{equation}
\begin{equation}
    \begin{split}
        \hat{S}^{(3)}_{\mathrm{RRV,}q\bar{q},b}=&\frac{1}{9\epsilon^4}+\frac{1}{54\epsilon^3}+\frac{1}{\epsilon^2}\left(-\frac{2}{9} \zeta _2-\frac{17}{81}\right)+\frac{1}{\epsilon}\left(\frac{17}{27} \zeta _2-\frac{131}{486}\right)+\frac{511}{81} \zeta _2-8 \zeta _3-\frac{5}{3} \zeta _4\\
        &+\frac{4369}{729}+\epsilon\left(-\frac{176}{9} \zeta _3 \zeta _2-\frac{23701}{243} \zeta _2+\frac{1354}{9} \zeta _3-\frac{737}{18} \zeta _4+\frac{152}{3} \zeta _5+\frac{102139}{4374}\right),
    \end{split}
\end{equation}
\begin{equation}\label{eq:srrvqqc}
    \begin{split}
        \hat{S}^{(3)}_{\mathrm{RRV,}q\bar{q},c}=&\frac{2}{9 \epsilon ^4}+\frac{19}{27 \epsilon ^3}+\frac{1}{\epsilon ^2}\left(\frac{173}{81}-\frac{4 \zeta _2}{3}\right)+\frac{1}{ \epsilon }\left(\frac{118}{9} \zeta _ 2-\frac{164}{9} \zeta _ 3+\frac{76}{243}\right)\\
        &-\frac{568}{27} \zeta _2+\frac{3590}{27} \zeta _3-\frac{298}{3} \zeta _4-\frac{11026}{729}\\
        &+\epsilon\left(\frac{1144}{3} \zeta _3 \zeta _2+\frac{15526}{81} \zeta _2-\frac{25400}{81} \zeta _3+\frac{6373}{9} \zeta _4-1284 \zeta _5-\frac{241319}{2187}\right).
    \end{split}
\end{equation}

To validate our calculation, several checks have been made. All the MIs are checked by independent numerical calculations. The highest $\epsilon^{-5}$-pole in $C_F C_A^2$ is validated by the strongly ordered limit calculation. Also we observed the  cancellation of the $C_F C_A^2$ $\epsilon^{-5}$-pole among RRR (c.f. eqs.~(35-38) in ref.~\cite{Baranowski:2022khd}), RRV and VVR (see Eq.~(\ref{eq:vvrcfca2})) as required by the maximal non-abelian exponentiation theorem. We note that in the direct calculation, the highest pole is obtained as a consequence of the non-trivial cancellation between higher spurious poles, therefore the correctness of the leading pole serves as a non-trivial check of our approach. Another strong check is on the $C_F^2 C_A$ term, where the direct brute-force computation completely agrees with the non-abelian exponentiation theorem which guarantees that the $C_F^2 C_A$ term can be obtained by convolution of the ${\cal O}(\alpha_s^2)$ RV correction and the ${\cal O}(\alpha_s)$ soft function in
Eq.~(\ref{eq:rvsoft}) and Eq.~(\ref{eq:losoft}), respectively. Further checks have been performed for the class of contributions, e.g. vacuum polarization of the gluon propagator, in which we can work out the virtual loop integration first and then evaluate the double real emissions analytically. In those checks, again we find the complete agreement with the computational setups using tensor reduction and MIs in this manuscript. 

\section{Conclusions}~\label{sec:end}
In this manuscript, we present the double-real-virtual (RRV) as well as the double-virtual-real (VVR) contributions to the three-loop thrust soft function. The main results of this work are presented in Eq.~(\ref{eq:srrvg}) through Eq.~(\ref{eq:srrvqqc}) for RRV and in the Appendix~\ref{appendix:vvr} for VVR, respectively. 

Our calculation paves the way for the complete N${}^3$LO thrust soft function, which is the key ingredient for the N${}^3$LL' and N${}^4$LL resummation of the thrust observable in the di-jet configuration. The RRV result can also be directly used for the 0-jettiness in the Drell-Yan/$ggH$ process as well as the $1$-jettiness observable in DIS, while the analytic continuation of the VVR to these processes is straightforward. Furthermore, the results presented here are the indispensable components to fulfill the N${}^3$LO calculation using the $N$-jettiness subtraction scheme~\cite{Boughezal:2015dva,Gaunt:2015pea,Caola:2022ayt}. 

The calculation demonstrates the feasibility of the recently proposed computational framework for a unified treatment of the loop and the phase space integration with Heaviside functions~\cite{Chen:2019fzm,Chen:2019mqc,Chen:2020wsh}. We note that beyond the thrust, the methodology used in this work can be used to other phase-space integrals
that involve experimental cuts through the Heaviside functions, and we particularly expect its applications to future jet and substructure precision calculations~\cite{Liu:2021xzi}. 

\textbf{Note}: In the earlier versions of this manuscript, we did not present the result for the fermionic contribution, which is added in this version. We note that a similar calculation was recently reported in ref.~\cite{Baranowski:2024ene}. Perfect agreement is found between these two independent calculations.

\acknowledgments
W.~C. is supported by the Natural Science Foundation of China (NSFC) under the contract No. 11975200.
F.~F. is supported by the National Natural Science Foundation of China under Grant No. 11875318, No. 11505285, and by the Yue Qi Young Scholar Project in CUMTB.  
Y.~J. is supported in part by the National Natural Science Foundation of China under Grants No. 11925506, 11875263, and No. 12070131001 (CRC110 by DFG and NSFC).
X.~L. is supported by the National Natural Science Foundation of China under Grant No.~12175016 and by the Guangdong Provincial Key Laboratory of Nuclear Science with No.2019B121203010 

\appendix 

\section{Double-Virtual-Real (VVR) Contributions}\label{appendix:vvr}

The calculation of the one-emission contribution to the thrust soft function is straightforward and is obtained by evaluating 
\bea\label{eq:vlr} 
S^{(l+1)}_{V^{l}R} (\tau;\epsilon)
= \int \frac{d^Dk }{(2\pi)^D}
(2\pi)\delta^+(k^2)  \, 
\omega_{V^{(l)}R}(k;n,{\bar n})\,
\Theta_\tau(k)
\,, 
\eea 
where $l$ denotes the number of the loops.
The phase space integral can be parameterized as
\bea 
\frac{d^D k}{(2\pi)^D}
2\pi \delta^+(k^2)
= 
\frac{d \Omega_{D-2}}{4(2\pi)^{D-1}}
dk^+ dk^- 
(k^+k^-)^{(D-4)/2}  \,, 
\eea 
where 
$\Omega_{D-2} = \int d\Omega_{D-2} = 
\frac{2\pi^{\frac{D-2}{2}}}{\Gamma((D-2)/2)}
$ is the solid angle. 
The matrix element $\omega_{V^{(l)}R}$ receives contributions
\bea\label{eq:Mvvr} 
\omega_{V^{(l)}R}
= {\cal J}_{V^{l}R}^{\dagger}
{\cal J}_R + c.c. 
+ \sum_{l_1=1}
{\cal J}_{V^{l-l_1}R}^{\dagger}
{\cal J}_{V^{l_1}R} \,,
\eea 
where ${\cal J}$ is the soft current. The tree level soft current is nothing but the well-known eikonal factor, 
\bea 
{\cal J}_R^{\mu,a}(k)  = g_s \mu^\epsilon \sum_{i=n,{\bar n}} T^a_i \frac{p^\mu_i}{p_i \cdot k} \,, 
\eea 
and the $1$-loop current is known to be~\cite{Catani:2000pi}  
\bea 
{\cal J}_{VR}^{\mu,a}(k)
= - g_s^3 \mu^{3\epsilon}\frac{S_\epsilon }{16\pi^2}
\frac{\Gamma(1-\epsilon)\Gamma(1+\epsilon)}{\epsilon^2}
i f^{abc} \sum_{i\ne j}T_i^b T_j^c 
\left( \frac{p_i^\mu}{p_i\cdot k } - \frac{p_j^\mu}{p_j\cdot k} 
\right) 
\left[ 
\frac{(-s_{ij})}{(-s_{ik})(-s_{jk})}
\right]^\epsilon  \,, \qquad 
\eea 
where  
\bea 
S_\epsilon = (4\pi)^\epsilon 
\frac{\Gamma(1+\epsilon)\Gamma^2(1-\epsilon)}{\Gamma(1-2\epsilon)} \,. 
\eea 

In the thrust case that we are studying, 
the first term in Eq.~(\ref{eq:Mvvr}) takes the general form~\cite{Duhr:2013msa} 
\bea 
{\cal J}_{V^{l}R}^{\dagger}
{\cal J}_R + c.c. 
= -  (4\pi\alpha_s)^{(l+1)}
4 C_i \left( 
\frac{S_\epsilon}{16\pi^2}
\right)^l
\left[
\frac{n\cdot \nbar 
\, e^{-i\lambda_{n\nbar} \pi} 
}{2 n\cdot k \nbar \cdot k 
\, e^{-i (\lambda_{nk} + \lambda_{\nbar k})\pi }
} 
\right]^{1+l\epsilon} 
\, 
r^{(l)}_{soft} + c.c. \,, 
\eea 
where $\alpha_s$ is the bare strong coupling constant and 
$C_i = C_F$ for $q {\bar q}$ process while $C_i = C_A$ for $gg$. 
$\lambda_{ij} = 1$ if both $i$ and $j$ are incoming or outgoing, otherwise $\lambda_{ij} = 0$. 
The factor $r_{soft}^{(l)}$ is given by~\cite{Duhr:2013msa} 
\bea 
r_{soft}^{(0)} = \frac{1}{2}\,, \qquad 
r_{soft}^{(1)} = - 
\frac{C_A}{\epsilon^2}
\Gamma(1-\epsilon)\Gamma(1+\epsilon) \,, 
\eea 
while 
\bea 
r_{soft}^{(2)}
= C_A N_f \,  r_1(\epsilon)
+ C_A^2  \, r_2(\epsilon) \,,
\eea 
with 
\bea 
&& r_1(\epsilon) 
= \frac{2\Gamma(-2\epsilon)}{(1+\epsilon)\Gamma(4-2\epsilon)} 
\frac{1
}{\cos^2(\pi \epsilon )}
\left[ 
3 \frac{\Gamma(1-\epsilon)
\Gamma(1-2\epsilon)}{\Gamma(1-3\epsilon)}
- \frac{1+\epsilon^3}{\epsilon^2(1+\epsilon)}
\frac{\Gamma^2(1-2\epsilon)}{\Gamma(1-4\epsilon)}
\right] \,,  \nn \\
&& r_2(\epsilon)
= \frac{\Gamma(1-\epsilon)
\Gamma(1-2\epsilon)
}{6 \epsilon^4
\cos^2(\pi\epsilon)\Gamma(1-3\epsilon) }
  \Bigg\{ 
(1+4\epsilon) 
{}_4F_3(1,1,1-\epsilon,-4\epsilon;2,1-3\epsilon,1-2\epsilon,1)  \nn \\ 
&&
-6\epsilon 
\left[ 
\psi(1-3\epsilon)
+ \psi(1-2\epsilon) 
- \psi(1-\epsilon) 
-\psi(1+\epsilon) 
\right] 
+ \frac{-3+5\epsilon+4\epsilon^2+14\epsilon^3}{2(1+\epsilon)(3-2\epsilon)(1-2\epsilon)}
\Bigg\} \nn \\
&& +  
\frac{1+4\epsilon}{3\epsilon^4(1+2\epsilon) }
\frac{\Gamma^2(1-2\epsilon)}{\cos^2(\pi \epsilon)\Gamma(1-4\epsilon)} 
\Bigg\{ 
2\, {}_3F_2\left(
1,-2\epsilon,1+2\epsilon;1-\epsilon,2+2\epsilon;1
\right) \nn \\ 
&& \hspace{10.ex}
- 
\frac{\Gamma(1+\epsilon)\Gamma(1-2\epsilon)}{\Gamma(1-\epsilon) } \, 
{}_3F_2(-2\epsilon,1+\epsilon,1+2\epsilon;1-\epsilon,2+2\epsilon;1) \nn \\
&& \hspace{10.ex}+ 
\frac{
(1+2\epsilon)
(3-38\epsilon
-16\epsilon^2 
+ 13\epsilon^3
+6\epsilon^4
)}{
4 
(1+\epsilon)
(1-2\epsilon)
(1+4\epsilon)
(3-2\epsilon) 
} \Bigg\} \,,
\eea 
where $\psi$ is the digamma function and ${}_pF_q$ are the generalized Hypergeometric functions.

The integration in Eq.~(\ref{eq:vlr}) can be evaluated straightforwardly which gives the VVR correction to the thrust soft function with renormalized strong coupling constant $\alpha_{s,\mathrm{r}}$
\bea 
S^{(3)}_{V^2R}(\tau) 
= \frac{\alpha_{s,\mathrm{r}}^3}{8\pi^3} 
\, 
\frac{1}{\mu}
\left(\frac{\tau}{\mu}\right)^{-1-6\epsilon} \,
\Big(
C_F C_A N_f S^{(3)}_{V^2R,C_FC_An_f}
+ 
C_F C_A^2 S^{(3)}_{V^2R,C_FC_A^2}
\Big) \,, 
\eea 
where 
\bea 
S^{(3)}_{V^2R,C_FC_An_f} 
&=& 
-\frac{\pi ^{5/2}  4^{\epsilon } e^{3 \gamma_E  \epsilon } \left(\cos^2(\pi  \epsilon )- \sin^2(\pi \epsilon )\right)}{3 
\Gamma (-2 \epsilon )
\Gamma \left(\frac{1}{2}-\epsilon \right)  \sin^2(\pi \epsilon)} \, r_1(\epsilon)
\nn \\
&=& \frac{1}{9 \epsilon^4}
+ \frac{5}{27\epsilon^3}
+ \frac{1}{\epsilon^2}
\left( 
\frac{19}{81}
-\frac{23}{18}\zeta _2
\right) 
+ \frac{1}{\epsilon}
\left( 
\frac{65}{243}
-\frac{115}{54}\zeta_2
-\frac{7}{3}  \zeta_3
\right)  \nn \\ 
&& 
+\frac{211}{729} 
-\frac{545}{162} \zeta_2
-\frac{35 }{9}\zeta_3
-\frac{13}{48} \zeta_4\nn\\
&&+ \left(
\frac{665}{2187}
-\frac{2359 \zeta_2}{486}
-\frac{223 \zeta_3}{27}
-\frac{65 \zeta_4}{144}
-\frac{341 \zeta_5}{15}
+ \frac{161 \zeta_3 \zeta _2}{6}
\right) \epsilon \,, \nn \\ 
\eea 
and 
\bea\label{eq:vvrcfca2} 
S^{(3)}_{V^2R,C_FC_A^2}
&=& 
-\frac{\pi ^{5/2}  4^{\epsilon } e^{3 \gamma_E  \epsilon } \left(\cos^2(\pi  \epsilon )- \sin^2(\pi \epsilon )\right)}{3 
\Gamma (-2 \epsilon )
\Gamma \left(\frac{1}{2}-\epsilon \right)  \sin^2(\pi \epsilon)} \, r_2(\epsilon) 
+ \frac{\pi ^4 e^{3 \gamma_E  \epsilon } \csc ^4(\pi  \epsilon ) \Gamma (-\epsilon )}{6 \epsilon ^2 \Gamma^2 (-2 \epsilon )}
\nn \\
&=& - \frac{1}{3\epsilon^5}
- \frac{11}{18 \epsilon^4}
- \frac{1}{\epsilon^3} 
\left( 
\frac{67}{54}
+ \frac{23}{6}\zeta_2 
\right) 
- \frac{1}{\epsilon^2}
\left( 
\frac{193}{81} -
\frac{253}{36}\zeta_2
- 2\zeta_3 
\right)  \nn \\ 
&& +\frac{1}{\epsilon} 
\left( 
- \frac{1142}{243}
+ \frac{1541}{108}\zeta_2
+ \frac{77}{6}\zeta_3
+ \frac{409}{48} \zeta_4
\right)  \nn \\
&&
-\frac{6820}{729}
+\frac{4547}{162} \zeta_2
+\frac{469 }{18} \zeta_3
+\frac{143 \zeta_4}{96}
+ \frac{52 }{3}
\zeta _2 \, \zeta_3 
-\frac{122 }{15} \zeta_5\,,
\nn \\
&&
+ 
\left( 
-\frac{40856}{2187}
+\frac{13403 \zeta _2}{243}
+\frac{1441 \zeta _3}{27}
+\frac{871 \zeta _4}{288}
+\frac{3751 \zeta _5}{30}
-\frac{1771 \zeta _2 \zeta _3}{12}
-\frac{637 \zeta _6}{1152}
-\frac{31}{2}  \zeta _3^2
\right) \epsilon  \,, \nn \\ 
\eea 
where the first term comes from the interference between the $2$-loop and the tree-level soft current and the second contribution from the square of the $1$-loop soft current. 

Here we have performed the
renormalization of the strong coupling constant $\alpha_s$, which is given by
\bea 
\alpha_s  \, 
\mu^{2\epsilon}  
\to \alpha_{s,\mathrm{r}}  
(4\pi)^{-\epsilon} e^{\gamma_E \epsilon}
\mu^{2\epsilon}
\left(  
1-
\frac{\alpha_{s,\mathrm{r}}}{2\pi}
\frac{\beta_0}{\epsilon}
+ \frac{\alpha_{s,\mathrm{r}}^2}{4\pi^2}
\left(
\frac{\beta_0^2}{\epsilon^2}
- 
\frac{\beta_1}{2\epsilon}
\right)
+ \dots 
\right) \,, 
\eea 
where $\alpha_{s,{\rm r}}$ is the renormalized strong coupling and $\gamma_E$ is the Euler constant.
The first $2$ orders of the $\beta$ function are given by
\bea 
&& \beta_0 = \frac{11}{6}C_A 
- \frac{1}{3} N_f \,,  \qquad
\beta_1 = \frac{17}{6} C_A^2
- \frac{5}{6} C_A N_f 
- \frac{1}{2} C_F N_f \,. 
\eea 

The renormalization of the strong coupling constant gives rise to the correction to the N$^3$LO soft function of the form
\bea 
S^{(3)}_{ren.R}(\tau) 
=
 \frac{\alpha_{s,\mathrm{r}}^2}{4\pi^2}
\left(
\frac{\beta_0^2}{\epsilon^2}
-
\frac{\beta_1}{2\epsilon}
\right)
S^{(1)}_R(\tau) 
- \frac{\alpha_{s,\mathrm{r}}}{2\pi} 
\frac{2\beta_0}{\epsilon} S^{(2)}_{VR}(\tau)  \,, 
\eea 
where $S^{(1)}_R(\tau)$ and $S_{VR}^{(2)}(\tau)$ are the NLO soft function and the real-virtual correction to the ${\cal O}(\alpha_s^2)$ soft function, respectively. They are found to be
\bea\label{eq:losoft} 
S_R^{(1)}(\tau)
= \frac{\alpha_{s,\mathrm{r}}}{2\pi} C_F
\frac{1}{\mu}
\left(\frac{\tau}{\mu} \right)^{-1-2\epsilon}
\frac{4 e^{\gamma_E \epsilon}}{\epsilon \, \Gamma(1-\epsilon) }
\,, 
\eea 
and 
\bea\label{eq:rvsoft}  
S_{VR}^{(2)}(\tau) 
= - \frac{\alpha_{s,\mathrm{r}}^2}{4\pi^2} C_F C_A
\frac{1}{\mu}
\left(\frac{\tau}{\mu} \right)^{-1-4\epsilon}
\, 
\frac{2 e^{2 \gamma_E \epsilon}}{\epsilon^3}
\frac{\Gamma^3(1-\epsilon)\Gamma^3(1+\epsilon)}{\Gamma^2(1-2\epsilon)\Gamma(1+2\epsilon)} 
\,. 
\eea

\section{Double-Real (RR) Contributions}\label{appendix:rr}
In this appendix, we present the ${\cal O}(\alpha_s^2)$ double-real (RR) correction (up to order $\mathcal{O}(\epsilon^2)$) for the thrust to validate our computational framework. The result is consistent with that in ref. ~\cite{Baranowski:2020xlp}.

The two-gluon contribution reads

\begin{equation}
    \begin{split}
       \hat{S}^{(2)}_{\text{RR,gluon}}=&(4\pi)^5C_F C_A\alpha_s^2\left\{\frac{1}{\epsilon ^3}+\frac{43}{24 \epsilon ^2}+\frac{1}{18\epsilon }[65-90 \zeta_2]+\frac{391}{54}-\frac{43 }{6}\zeta_2-9 \zeta_3\right.\\
       &\left.+\epsilon  \left[\frac{133}{9} \zeta_2-\frac{215}{6}\zeta_3-\frac{21 }{2}\zeta_4+\frac{994}{81}\right]\right.\\
       &\left.\epsilon^2\left[30 \zeta_3 \zeta_2-\frac{752}{27} \zeta_2+\frac{397}{3} \zeta_3-\frac{344}{3} \zeta_4-91 \zeta_5+\frac{5588}{243}\right]\right\}\\
       &+C_F^2 \left\{-\frac{4}{\epsilon ^3}+\frac{24}{\epsilon } \zeta_2+64 \zeta_3+48 \zeta_4\epsilon+\epsilon^2\left[768 \zeta_5-384 \zeta_2 \zeta_3\right]\right\}.
    \end{split}
\end{equation}

\noindent The two-ghost contribution reads

\begin{equation}
    \begin{split}
    \hat{S}^{(2)}_{\text{RR,ghost}}=&(4\pi)^5C_F C_A\alpha_s^2\left\{\frac{1}{24 \epsilon ^2}+\frac{1}{9 \epsilon }+\frac{13}{54}-\frac{\zeta_2}{6}+\frac{\epsilon}{81} \left[9\zeta_2-\frac{135}{2}\zeta_3+76\right]\right.\\
    &\left.+\frac{\epsilon^2}{27}\left[-20 \zeta_2+45 \zeta_3-72 \zeta_4+\frac{620}{9}\right]\right\}.
    \end{split}
\end{equation}

\noindent The two-quark contribution reads

\begin{equation}
    \begin{split}
    \hat{S}^{(2)}_{\text{RR,quark}}=&(4\pi)^5C_F N_f\alpha_s^2\left\{
    -\frac{1}{3 \epsilon ^2}
    -\frac{5}{9 \epsilon }+
    \frac{4 \zeta_2}{3}-\frac{28}{27}-\epsilon  \left[\frac{32}{9} \zeta_2-\frac{20 }{3}\zeta_3+\frac{20}{81}\right]\right.\\
    &\left.+\frac{\epsilon^2}{27}\left[136 \zeta_2-792 \zeta_3+576 \zeta_4+\frac{536}{9}\right]\right\}.
    \end{split}
\end{equation}

\bibliographystyle{JHEP}
\bibliography{references}

\end{document}